\documentclass{ieeeaccess}

\usepackage{cite}
\usepackage{amsmath,amssymb,amsfonts}
\usepackage{algorithmic}
\usepackage{graphicx}
\usepackage{textcomp}

 \usepackage{comment}
 \usepackage{makecell}
 \usepackage{booktabs}

\usepackage{enumitem}

\usepackage{siunitx}
\usepackage[dvipsnames]{xcolor}

\usepackage[hyphens]{url}

\usepackage{microtype}
\usepackage{balance}
\usepackage{hyperref}

\newcommand{\orcid}[1]{\href{https://orcid.org/#1}{\includegraphics[width=8pt]{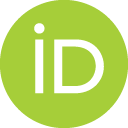}}}
\usepackage{draftwatermark}
\SetWatermarkText{This is the pre-print\\version}
\SetWatermarkColor[gray]{0.75}
\SetWatermarkFontSize{2cm}
\SetWatermarkAngle{45}
\SetWatermarkHorCenter{10cm}

\begin{document}
	\history{Date of publication xxxx 00, 0000, date of current version xxxx 00, 0000.}
	\doi{10.1109/ACCESS.2017.DOI}
	
	\title{A Taxonomy for Quality in Simulation-based Development and Testing of Automated Driving Systems}
	\author{\uppercase{Barbara Schütt \orcid{0000-0001-8439-0322}}\authorrefmark{1},
		\uppercase{Markus Steimle \orcid{0000-0001-8913-6980}\authorrefmark{2}, Birte Kramer \orcid{0000-0002-9982-0206}\authorrefmark{3}, Danny Behnecke \orcid{0000-0001-5228-5640}\authorrefmark{4}, and Eric Sax\authorrefmark{1}}}
	\address[1]{FZI Research Center for Information Technology, Karlsruhe, Germany (e-mail: schuett,sax@fzi.de)}
	\address[2]{TU Braunschweig, Braunschweig, Germany (e-mail: steimle@ifr.ing.tu-bs.de)}
	\address[3]{OFFIS e.V., Oldenburg, Germany (e-mail: birte.kramer@offis.de)}
	\address[4]{German Aerospace Center, Braunschweig, Germany  (e-mail: Danny.Behnecke@dlr.de)}
	\tfootnote{This research is funded by the SET Level research initiative, promoted by the German Federal Ministry for Economic Affairs and Energy (BMWi)}
		
	\markboth
	{Schütt \headeretal: A Taxonomy for Quality in Simulation-based Development and Testing of Automated Driving Systems}
	{Schütt \headeretal: A Taxonomy for Quality in Simulation-based Development and Testing of Automated Driving Systems}
	
	\corresp{Corresponding author: Barbara Schütt (e-mail: schuett@fzi.de).}
	
	\begin{abstract}
	Ensuring the quality of automated driving systems is a major challenge the automotive industry is facing. 
	In this context, quality defines the degree to which an object meets expectations and requirements.
	Especially, automated vehicles at SAE level 4 and 5 will be expected to operate safely in various contexts and complex situations without misconduct. 
	Thus, a systematic approach is needed to show their safe operation.
	A way to address this challenge is simulation-based testing as pure physical testing is not feasible.
	During simulation-based testing, the data used to evaluate the actual quality of an automated driving system are generated using a simulation. 
	However, to rely on these simulation data, the overall simulation, which also includes its simulation models, must provide a certain quality level.
	This quality level depends on the intended purpose for which the generated simulation data should be used. 
	Therefore, three categories of quality can be considered: quality of the automated driving system and simulation quality, consisting of simulation model quality and scenario quality.
	Hence, quality must be determined and evaluated in various process steps in developing and testing automated driving systems, the overall simulation, and the simulation models used for the simulation.
	In this paper, we propose a taxonomy to serve a better understanding of the concept of quality in the development and testing process to have a clear separation and insight where further testing is needed -- both in terms of automated driving systems and simulation, including their simulation models and scenarios used for testing.
	\end{abstract}
	
	
	\titlepgskip=-15pt
	
	\maketitle
	
	\section{Introduction}
\label{sec:introduction}
Future mobility systems will face many challenges as the growing urbanization may bring transportation systems to their limits \cite{Urbanization}. 
Thus, problems we are already facing today, e.g., traffic jams or accidents, will likely become worse.  
Assisted and automated driving systems have the potential to meet these challenges by promising more mobility for everyone, driving more efficiently, environmentally friendly, and safely.
This leads to growing system complexity and interconnection of automotive features \cite{bach2017taxonomy}.
Above all, safety will be a crucial issue for the introduction and acceptance of these systems in society. 
Thus, it is essential to ensure and validate the vehicles' safe behavior.
For this purpose, the vehicle must be thoroughly tested at various test levels 
to assure requirements are met, the system has necessary capabilities in all intended use cases, and unreasonable risk is avoided \cite[p.~6]{wood_safety_2019}. 

Currently, real-world test drives are used to validate the safe behavior of vehicles equipped with assisted driving systems.
These test results give a statement about their quality, ``the degree of excellence'' \cite{dict2020}.
According to Wachenfeld and Winner \cite{wachenfeld2016release}, the distance-based test approach is no longer feasible at a certain degree of automation since more than six billion test kilometers would be necessary to ensure the vehicles' safe behavior.
Additionally, if there are changes or variations in the automated driving system, all testing has to be repeated.
Scenario-based test approaches promise an alternative or supplemental test method, especially combined with a simulation-based approach. 
Compared to the random scenarios emerging during a real-world test drive, in scenario-based testing, new and relevant scenarios are systematically derived and tested at different process steps during the development and test process \cite{pegasus_method}, \cite{menzel2018scenarios}.
These approaches aim to create a collection of relevant scenarios, depending on the test object, the test objectives, and the test object's preceding requirements and can be executed in simulation-based or real-world test, e.g., vehicle in loop.
Simulation-based tests are significantly cheaper than real-world test. 
With that in mind, it has to be ensured that the system under test (SUT) meets its specifications, but additionally, all used simulation models and tools must possess a certain quality level to generate sufficiently valid simulation results \cite{Bode}.
Therefore, three categories of quality can be distinguished: the quality of the automated vehicle and the quality of the simulation, consisting of simulation models as the simulation environment and scenarios as the schedule of events.
The system under test depends on the category and test level considered.
Within the first category, the system under test can be, for example, the overall automated vehicle or a system, component, or a unit of the automated vehicle like the automated driving system or a path planning algorithm.
Within the second and third categories, the system under test can be, for example, the overall simulation, coupled simulation models, a single simulation model, or a scenario.
All categories have to be considered when simulation-based approaches are used.

While the proposed approach might me applicable for other areas of safety-critical simulation, e.g., flight simulation, it mainly focuses on automotive simulation and its evaluation.
However, the generalized parts of our work, e.g. terminology, may be applicable to other domains. This comparison between domains was not focus of this work but might be considered in future work.

\subsection*{Novelty and main contribution to the state of the art} 

 The novelty and main contribution of this paper is a taxonomy for a systematical classification of quality during the simulation in the development and test process of automotive applications. 
 Therefore, we

\begin{itemize}
 \item give a definition of relevant quality-related terms to avoid ambiguity,
 \item divide the simulation quality into three main categories that may occur during scenario-based testing and simulation: the quality regarding the simulation model or tool, the quality of a vehicle or components of it, and the quality of scenarios for testing, and
 \item propose an expandable taxonomy and terminology for all three categories and levels of resolution, to serve a categorization of quality metrics and aspects and to make it easier to communicate the area of interest within a simulation setup. The taxonomy uses the classification as a way to help scientists imminently understand and organize the differences of all areas of quality in the field of automotive simulation.
\end{itemize}

\subsection*{Structure}

In section~\ref{sec:rel_work}, scenario-based testing and different quality metrics are described.
Section~\ref{sec:glossary} defines relevant quality-related terms.
Section~\ref{sec:tax_complete} introduces the proposed taxonomy for quality throughout simulation-based testing and gives a simulation example that shows that each category and resolution level plays a role during the complete development and test process. 
Additionally, examples from literature to show the taxonomy’s role in current research were collected.
Finally, section~\ref{sec:conclusion} gives a short conclusion.

	\section{Related Work}
\label{sec:rel_work}
In this section, we summarize scenario-based testing, give an introduction to traffic simulation abstraction levels and quality metrics.

\subsection{Simulation- and scenario-based testing}

\Figure[t!](topskip=0pt, botskip=0pt, midskip=0pt)[width=0.99\columnwidth]{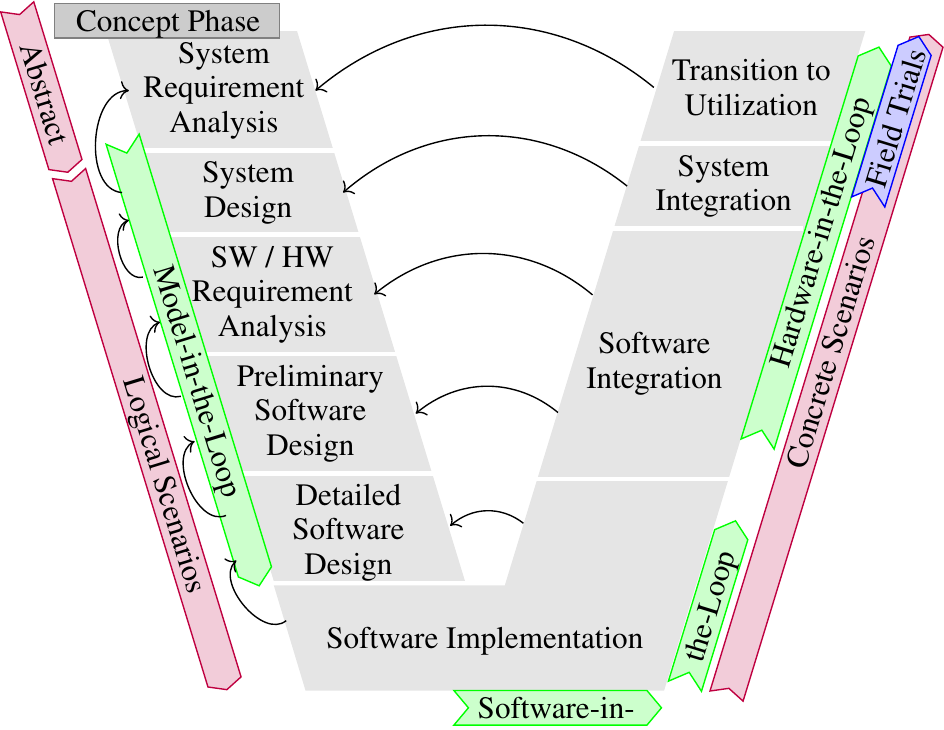}
{V-model with different X-in-the-loop phases \cite{king2019automated} combined with abstraction levels of scenarios at different development stages, adapted from Bock \textit{et al.} \cite{bock2019advantageous}.\label{fig:v_model}}


Quality assurance is an essential part of the development process.
Therefore, it shall be integrated into the development process as early as possible.
An established development approval is visualized by the V-model \cite{king2019automated} and shown in Fig.~\ref{fig:v_model}.
The left part of the V describes the top-down design process that comes after the concept phase.
The right part describes a bottom-up test process that includes verification and validation activities.
This process can be used for the development of a simulation model, the complete simulation tool, vehicle components, or the complete vehicle.
Depending on the proportion of simulated and real elements, the corresponding test method can be called, for example, HiL, SiL, VeHiL \cite{wachenfeld2016release}.

As stated by Wood \textit{et al.}\cite[pp.~83]{wood_safety_2019}, scenario-based testing is a suitable approach of supplementing the distance-based approach of real-world driving and thus reducing the necessary mileage.
The scenario-based approach includes the techniques and strategies during the test process listed below to gain information and make statements about the quality of a system under test:
\begin{itemize}
	\item decomposing the system and individual testing of system elements,
	\item combining different platforms and design techniques (e.g., variable variation or stochastic variation for increasing test coverage),
	\item defining scenarios by using expert knowledge about interesting situations or automatic extraction of traffic data, and
	\item defining surrogate metrics (e.g., crash potential of a given situation) and performance measures.
\end{itemize}

Based on Ulbrich \textit{et al.} \cite{ulbrich2015defining}, a scenario is defined as the temporal development of an initial scene and its participating static and dynamic actors, similar to a storyline.
According to Bach \textit{et al.} \cite{bach2016model}, scenarios can be divided into movie-related acts and use abstract propositional and temporal descriptions.
Consistency checks can be utilized to generate derivations of these scenarios to create a database with a collection of scenarios.
This movie-related storyboard approach is also taken up by OpenSCENARIO, an emerging scenario description standard \cite{openscenario}.

Menzel \textit{et al.} \cite{menzel2018scenarios} suggest three abstraction levels for scenarios: functional, logical, and concrete scenarios. 
Scenarios are developed at an abstract level during the concept phase \cite{kramer2020} and get detailed and concretized throughout the development and test process \cite{menzel2018scenarios}. 
According to Menzel \textit{et al.} \cite{menzel2018scenarios}, the most abstract level of scenario representation is called \textit{functional} and describes a scenario via linguistic notation using natural, non-structured language terminology.
The main goal for this level is to create scenarios easily understandable and open for discussion.
Bock \textit{et al.} \cite{bock2019advantageous} propose a supplementary abstraction level, called abstract scenarios, defined by a controlled natural language format that is machine-interpretable and an exemplary assignment of scenario abstraction levels to the V-model is shown in Fig.~\ref{fig:v_model}.
The next abstraction level is the \textit{logical} level and refines the representation of functional scenarios with the help of parameters. 
The most detailed level is called \textit{concrete}. 
It describes operating scenarios with concrete values for each parameter in the parameter space.
This means that one logical scenario can yield many concrete scenarios, depending on the number of variables, size of the range, and their step size.
A term related to scenario description is \textit{Operational Design Domain (ODD)}.
According to SAE \cite{sae2018surface}, the ODD defines conditions under which an automated driving system or feature is intended to function.
It determines \textit{where} (e.g., environmental or geographical characteristics) and \textit{when} (e.g., time-of-day restrictions) an automated driving system has to be able to act reliable.
Scenarios can be defined by scenario description languages, e.g., OpenSCENARIO~\cite{openscenario} or SceML~\cite{schutt2020sceml}.
Another related term is \emph{test case}.
According to Steimle \textit{et al.} \cite{Steimle2020TerminologyEnglish}, in scenario-based test approaches, a test case consists of at least a (concrete) scenario and evaluation criteria.

The automotive domain has brought forth a multitude of simulation software.
The following tools are currently relevant for this work, however, we do not warrant its completeness.
Commercial tools for automotive simulation are available from Vires VTD \cite{VTDsim}, dSpace \cite{dSpacesim},  and IPG \cite{IPGsim}.
All three simulation tools provide modules for map and scenario creation, sensor and dynamic models just to name some examples. 
A further example tool is Carla, an open-source simulator with a growing community and based on the game engine Unreal \cite{Dosovitskiy17}.
It offers several additional modules, e.g., a scenario tool which includes its own scenario format as well as support for OpenSCENARIO, a graphical tool for creating scenarios, a ROS-bridge, and SUMO support.
SUMO is a open source software tool for modeling microscopic traffic simulation from DLR \cite{SUMO2018}.
It specializes on big scale of traffic simulation and can be used for evaluating traffic lights cycles, evaluation of emissions (noise, pollutants), traffic forecast, and many others.
Other tools worth mentioning are: openPASS \cite{openPass}, PTV Vissim \cite{Vissim} or esmini \cite{esmini}, an OpenSCENARIO player.

\subsection{Simulation process}
The IEEE Std 1730-2010 \textit{IEEE Recommended Practice for Distributed Simulation Engineering and Execution Process (DSEEP)} \cite{5706287} defines processes and procedures to develop and execute distributed simulations.
It refers distributed simulation engineering and many automotive simulation tool and XiL-applications lie within its field.
It states seven main steps for simulation engineering:
\begin{itemize}
    \item[1] Define simulation environment objectives: Define and document the problem space that is addressed with the simulation environment.
    \item[2] Perform conceptual analysis: Developing a real-world representation regarding its problem space and scenario, requirement specification.
    \item[3] Design simulation environment: Select and design member applications and models.
    \item[4] Develop simulation environment: Implement member applications, models, and coupling methods.
    \item[5] Integrate and test simulation environment: Integrate member applications and test simulation environment.
    \item[6] Execute simulation: Execute planned simulation and document execution problems.
    \item[7] Analyze data and evaluate results: Analyze how well requirements are met and test criteria are fulfilled.
\end{itemize}
In particular step five to seven are relevant for this work since simulation and simulation model quality fall into the scope of DSEEP.

Furthermore, Durak \textit{et al.} \cite{durak2020safety} propose a simulation qualification level (SQL) that states how strictly a simulation environment in sense of reliability and credibility is evaluated. Their idea concerns flight simulation, however, translates to the field of automotive simulation and reminds of the tool confidence level of ISO 26262 \cite{iso26262}.

\subsection{Abstraction levels of traffic simulation}
\label{sec_rw_levels}
In general, traffic simulation can be divided into different abstraction levels regarding the depth of resolution: nanoscopic, microscopic, mesoscopic, and macroscopic \cite{ni2006framework}, \cite{schiller2019multi}, \cite{kups1274} as shown in Fig.~\ref{fig:sim_resolution_tikz}. 
In macroscopic traffic simulation, the traffic is modeled as fluid and is used to evaluate traffic flows or congestion in heavy traffic situations.
The next resolution level is mesoscopic traffic simulation.
Every participant is modeled as a discrete particle with its position but still lacks personality, such as mass or size.
This lacking personal information is added at the microscopic level.
At this level of resolution, each participant has its own modeled behavior with an individual state and variables, such as mass, speed, and acceleration.
Additionally, individual maneuvers relevant for specific scenarios are modeled.
An example for microscopic traffic simulation is the SUMO framework \cite{SUMO2018}.
The next abstraction level in traffic simulation is nanoscopic (sometimes called submicroscopic) and views each vehicle as a composition of different subunits that need to be coupled to achieve a higher level of detail.
Scenario-based testing often occurs in microscopic and nanoscopic simulation since the main goal is to evaluate (sub)units and their individual behavior in given scenarios.

Further examples can be found in section~\ref{sec_rw_metrics}.

\Figure[t!](topskip=0pt, botskip=0pt, midskip=0pt)[]{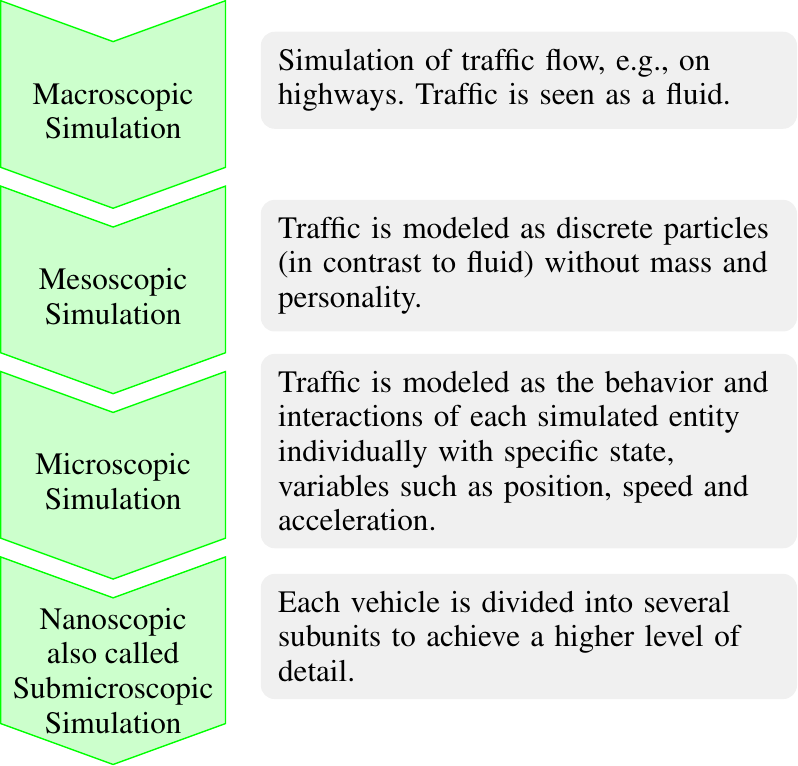}
{Different levels of resolution in traffic simulation \cite{ni2006framework}, \cite{schiller2019multi}.\label{fig:sim_resolution_tikz}}


\subsection{Quality metrics}
\label{sec_rw_metrics}
To define suitable metrics it must first be determined what needs to be tested, i.e. what is the system under test and its requirements, and which aspects should be considered in the corresponding test cases \cite{pegasus_method}.
In a less formal phase, e.g., prototyping a proof-of-concept for a simulation model, at least goals for the intended functionality have to be known.
From these requirements or goals, quality metrics can be derived and are essential parts of a test case \cite{Steimle2020TerminologyEnglish} to determine and quantify the quality of the intended functionality.

However, before assessing the system under test using data generated by simulation, the quality of the overall simulation and its individual simulation models must be assessed and ensured, e.g., tool qualification \cite{iso26262}. 
Viehof and Winner \cite{viehof2018research} introduced a method for objective quality assessment of simulation models by statistical validation, where a simulation model and its parameterization are validated separately.
This method has already been used successfully for vehicle dynamics simulation models and has been adapted for sensor perception simulation models by Rosenberger \textit{et al.} \cite{rosenberger2019towards}.
Furthermore, Riedmaier \textit{et al.} \cite{riedmaier2020framework} present a unified framework and survey for simulation model verification, validation, and uncertainty quantification. 
Nevertheless, to assess a simulation's quality, pure simulation model validity is not enough.
Through coupling and execution of the simulation models, more challenges have to be faced.
To our knowledge, there are no established or widely recognized verification or validation methods for simulation models and their coupling mechanisms.

Metrics to evaluate driver behavior or driving functions are more common, and there exists a long list of different possibilities \cite{westhofen2021criticality}.
Well-known metrics are surrogate safety measures to analyze the conflict potential or severity of microscopic traffic scenarios \cite{gettman2003surrogate}.
Example metrics are the calculation of the time-to-collision (TTC), post-encroachment time (PET), and Gap Time (GT).

The evaluation of scenario quality depends on the aspects of a scenario that are important for further test cases and scenarios.
Abeysirigoonawardena \textit{et al.} \cite{abeysirigoonawardena2019generating} use a distance-based metric with respect to two traffic participants for finding new scenarios, whereas Hallerbach \textit{et al.} \cite{hallerbach2018simulation} consider the current traffic situation, e.g., traffic density, to make a statement about a highway scenario.
Junietz \cite{junietz2019microscopic} proposes a metric to evaluate the macroscopic accident risk, the average risk of occurrence of fatal accidents, and the microscopic risk, describing the accident risk in a single scenario.

	\section{Terms and Definitions}
\label{sec:glossary}

In this section, terms related to quality aspects that are relevant for this paper are explained.
Fig.~\ref{fig:definitions} shows these terms and their relationships as a UML diagram.
Additionally, an example is given for each term (green boxes).
In the following paragraphs these terms are described.

\Figure[t!](topskip=0pt, botskip=0pt, midskip=0pt)[]{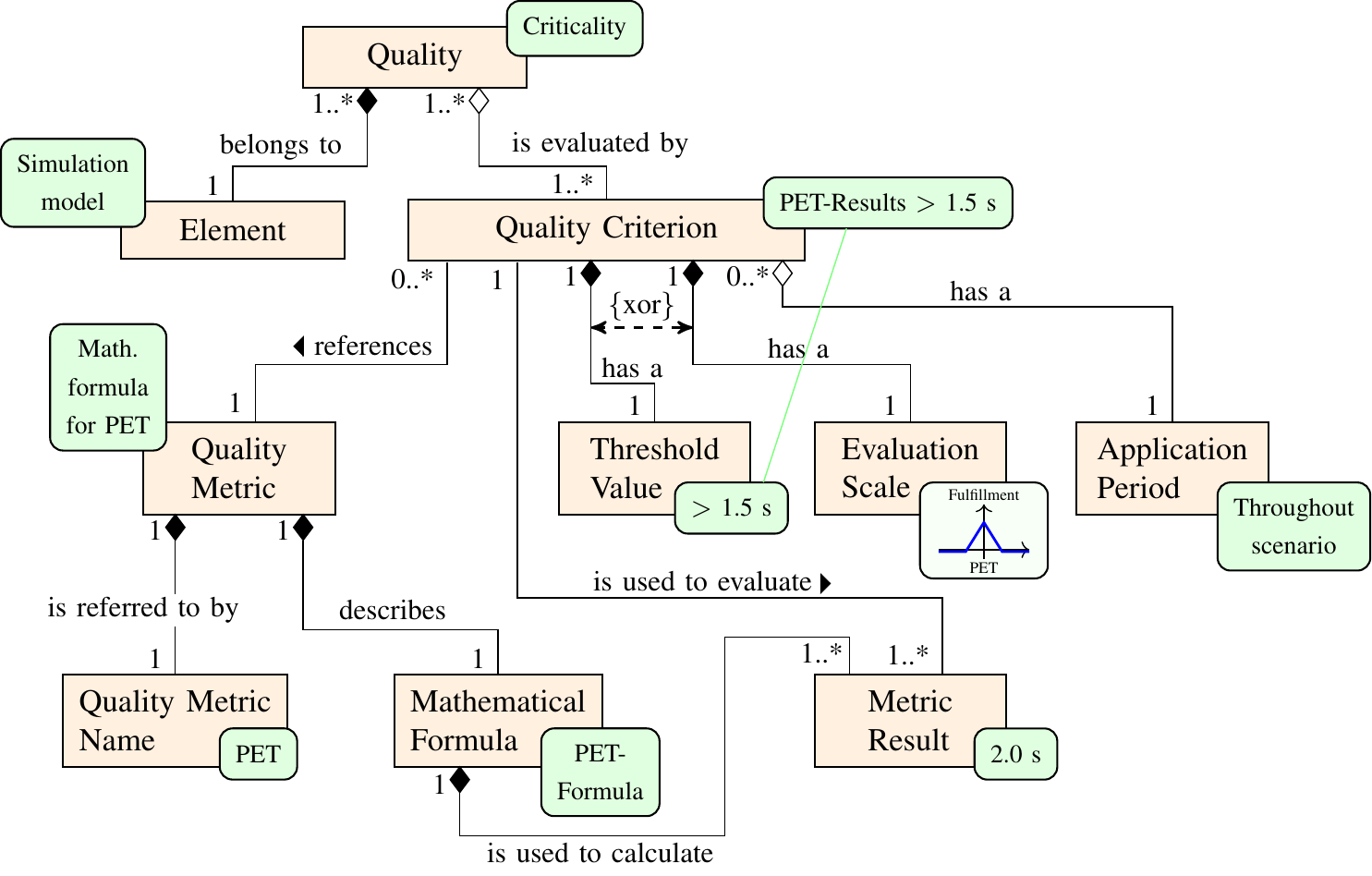}
{Relationship between relevant terms related to quality aspects. An example for each term is given in the assigned green boxes (PET: post-encroachment-time), based on \cite{Steimle2020TerminologyEnglish}.\label{fig:definitions}}

According to the Cambridge Dictionary~\cite{dict2020}, 
\textbf{Quality} is (1) ``the degree of excellence of something, often a high degree of it.''
Hence, quality always belongs to something we call element in this paper.
An \textbf{element} may be, e.g., a simulation or an automated driving function.
This quantified quality can be set in relation to other quantified qualities.
Quality is evaluated by one or more quality criteria.
This work distinguishes between different quality categories within the domain of automated driving, which are further explained in section~\ref{sec:tax_complete}.

The following terms and their descriptions are cited from Steimle \textit{et al.} \cite{Steimle2020TerminologyEnglish}. In the original text, Steimle~\textit{et~al.}~\cite{Steimle2020TerminologyEnglish} use the more general term ``evaluation.'' Since these descriptions refers to a quality, the term ``evaluation'' is replaced by ``quality.'' Additionally, we highlighted (in bold) the terms shown in Fig.~\ref{fig:definitions}.

``[A \textbf{quality}] \textbf{criterion} is used to evaluate one or more metric results in relation to a threshold value or an evaluation scale within a specified application period.
These metric results are calculated using a mathematical formula (described by [a quality] metric) and data generated during test case execution [...].
Thus, [a quality] criterion references [a quality] metric, it has a threshold value or an evaluation scale, and it has an application period.''~\cite{Steimle2020TerminologyEnglish}
An example quality criterion is the evaluation of post-encroachment-time calculated for a scenario execution and comparing to a threshold, e.g., a scenario is considered non-critical if post-encroachment-time exceeds $\SI{1.5}{\second}$ \cite{Allen.1978}.

``[A \textbf{quality}] \textbf{metric} is referred to by [a quality] metric name and describes a mathematical formula.
This formula is used to calculate one or more metric results based on data generated during test case execution [...].
Examples of [quality] metrics related to automated driving are the metric named time-to-collision (TTC) and the metric named post-encroachment-time (PET) (each including the associated mathematical formula).''~\cite{Steimle2020TerminologyEnglish}

``[A \textbf{quality}] \textbf{metric name} (e.g., time-to-collision (TTC) or post-encroachment-time (PET)) refers to a specific [quality] metric used to calculate one or more associated metric results.''~\cite{Steimle2020TerminologyEnglish}

``A \textbf{mathematical formula} (described by [a quality] metric) is a calculation rule used to convert input values (generated during test case execution) at a specific point in time into an output value (called metric result) that can be used for test case evaluation.''~\cite{Steimle2020TerminologyEnglish}
Post-encroachment-time uses the formula $PET = t_2 - t_1$, where $t_1$ denotes the time when actor 1 leaves the designated encroachment area and $t_2$ the time where actor 2 enters this area. Thus, post-encroachment-time is the distance in time between two actors passing the same area of interest.

``A \textbf{metric result} is calculated using a mathematical formula (described by [a quality] metric) and data [...] generated during test case execution [...].
A metric result is calculated at a certain point in time and consists of a number and a unit.
The calculated metric results are evaluated according to the corresponding [quality] criteria.''~\cite{Steimle2020TerminologyEnglish}
A possible PET result could be $\SI{3}{\second}$, which means actor 1 leaves the area of interest $\SI{3}{\second}$ before actor 2 enters it.

The metric results can be evaluated using two different methods, which usually exclude each other: using a threshold value or an evaluation scale.

``A \textbf{threshold value} is a fixed number (with a unit) used to test the compliance of calculated metric results with this fixed number according to the [quality] criterion.
Therefore, only a statement is possible regarding whether the [quality] criterion is fulfilled or not.''~\cite{Steimle2020TerminologyEnglish}

Allen \textit{et al.}\cite{Allen.1978} propose $\SI{1.5}{\second}$ as threshold for post-encroachment-time, where the result can be seen as critical if it falls below $\SI{1.5}{\second}$.

``An \textbf{evaluation scale} is a scale used to evaluate the adherence of calculated metric results with this scale according to the [quality] criterion.
Therefore, it is also possible to make a statement about how well the [quality] criterion is fulfilled.''~\cite{Steimle2020TerminologyEnglish}
Another interpretation of post-encroachment-time could include the severity of the potential critical situation that can be coupled to the distance in time and results in more severe situations when the time interval between two actors gets smaller.

``An \textbf{application period} defines the periods in which the corresponding [quality] criterion is applied. 
The application period is defined by one or more conditions that are linked with AND and/or OR operators. 
When the linked conditions are fulfilled, the application of the [quality] criterion starts. 
Its application continues until the linked conditions are no longer fulfilled, a specified time has elapsed, or a specified event has occurred.''~\cite{Steimle2020TerminologyEnglish}
A possible application period for post-encroachment-time is throughout the complete scenario.

A wide spread term in the automotive area is criticality or criticality metric.
From our point of view, (low) criticality is a subcategory of quality.
Therefore, when we mention quality criteria and quality metrics, they also include criticality criteria and criticality metrics.
	\section{Taxonomy for Simulation Quality}
\label{sec:tax_complete}


\Figure[t!](topskip=0pt, botskip=0pt, midskip=0pt)[width=0.88\textwidth]{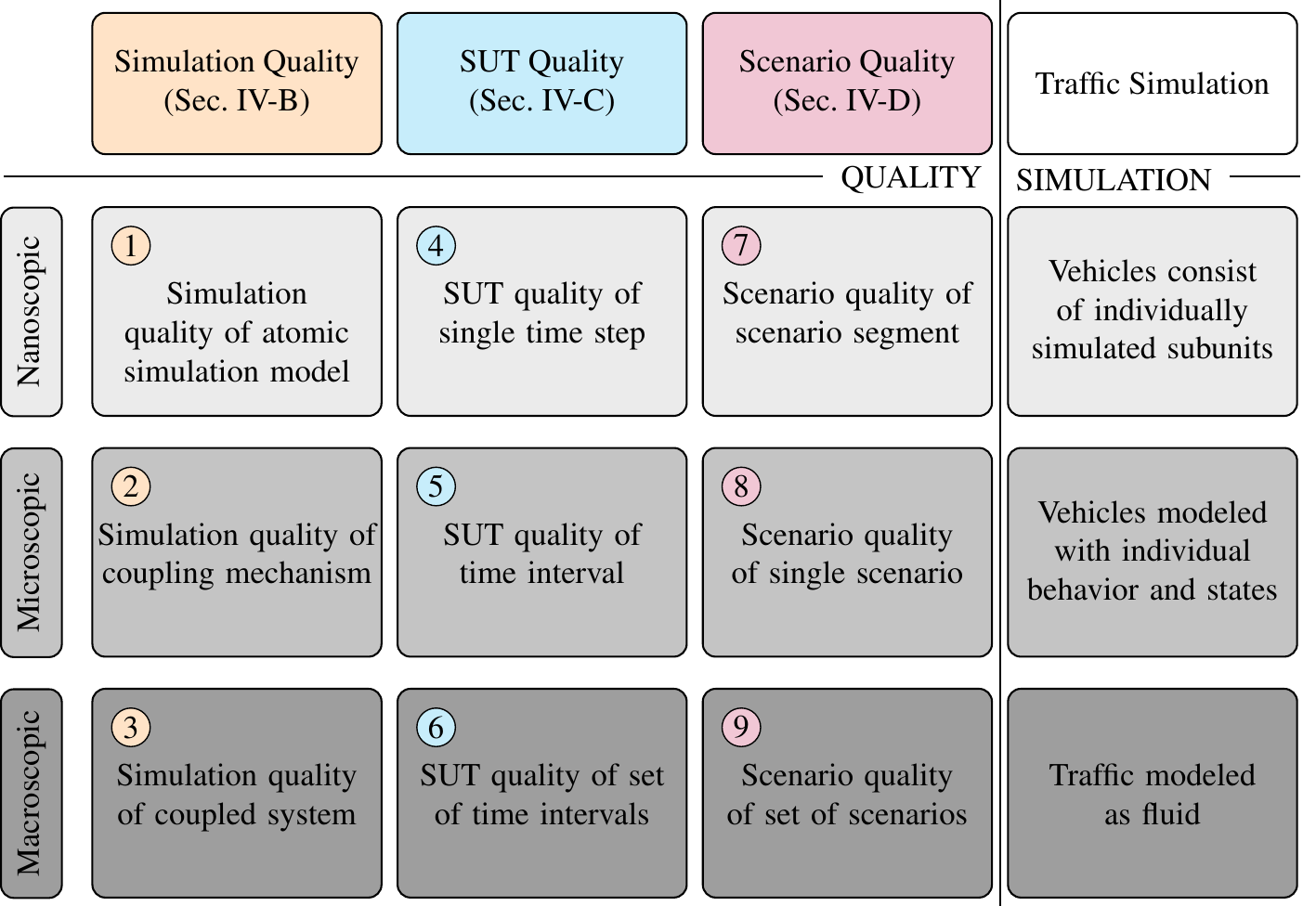}
{Quality matrix with three proposed quality categories and resolution levels: simulation quality, system under test (SUT) quality, scenario quality. The category of traffic simulation serves as entry point for this presentation.\label{fig:qualitymatrix}}

This section explains our taxonomy with its domains of interest and resolution levels in simulation quality.
In the last column it also shows the established resolution levels of traffic simulation as entry point.
In addition to traffic resolution, we propose three domains of interest for simulation quality (columns): simulation quality, system under test quality, and scenario quality.
All domains can be examined in different levels of resolution, as it is possible for traffic simulation (rows): nanoscopic, microscopic, and macroscopic.
In this work, we left out mesoscopic since it is not as common as the other three levels and does not add value to the taxonomy in general.
New rows, columns, or single entries can be added to our taxonomy if another resolution level or domain of interest is needed.
Further, this section states the role of all domains and their different resolutions during the development and test process.
Fig.~\ref{fig:qualitymatrix} shows all proposed combinations of domains of interest and resolution levels in a table, whereas Fig.~\ref{fig:chapter_flow} depicts the base structure of quality interaction.
The bracketed numbers in the text match those in the illustrations and show to which entry they belonged.

\Figure[t!](topskip=0pt, botskip=0pt, midskip=0pt)[]{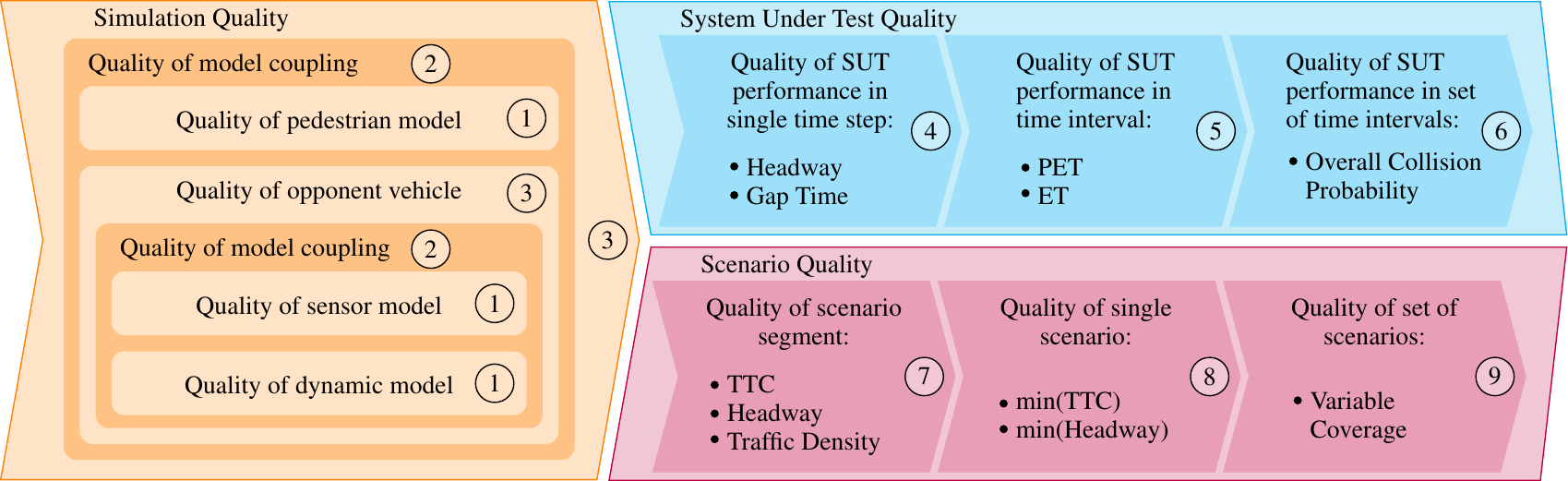}
{Basic structure of quality interaction described by section~\ref{sec:tax_complete} with examples. Abbreviated metrics are post-encroachment time (PET), encroachment-time (ET), time-to-collision (TTC).\label{fig:chapter_flow}}
\subsection{Simulation example}
\label{sec_example}

In the following description of the taxonomy, we use a simplified example for describing each domain of interest and its resolution levels:
\begin{quote}
	A highly automated driving function to avoid collisions at urban intersections without traffic signs or lights shall be tested with simulation-based methods. 
	Therefore, the driving function is the system under test.
	Its input is an object list that is available after sensor fusion. 
	The output is a target trajectory with additional information about target speed and acceleration for several time steps in the future. 
	A simulation environment shall be used for evaluation.
\end{quote}
An ODD for the example function in accordance with the definition of ODD from SAE \cite{sae2018surface} is defined:
\textit{The driving function is designed to operate at an urban intersection at daylight and at speeds not to exceed \SI{58}{\kilo\meter\per\hour}.}
Additionally to the ego vehicle, other traffic participants, e.g, pedestrians or opponent vehicles, are needed to test collision avoidance.
For our example scenarios we used dSpace Motion Desk \cite{dSpacesim} as simulation tool.

\subsection{Simulation Quality}
\label{sec_simmodel_q}
Simulation quality refers to the quality a simulation model has or is presumed to have.
This means the simulation model or their couplings are the elements from which the qualities are quantified.
Before a model can be used, it must be ensured that it approximates its real-world equivalent or functionality sufficiently in the relevant aspects.
Since we assume that our example function is safety-relevant, e.g., assigned automotive safety integrity level (ASIL) in accordance with ISO 26262 \cite{iso26262}, and our goal in the initial step is to test the function in a simulation-based approach, we must first determine the quality of the simulation tools in use.

According to Balci \cite{balci1998verification}, the process of ascertaining the simulation model quality consists of two parts: model verification and model validation.
Model verification evaluates the accuracy of transforming a problem formulation into a model, i.e., building the simulation model \textit{right}.
Model validation checks whether the simulation model is sufficiently valid for its intended purpose, i.e., building the \textit{right} simulation model.
Concerning verification and validation, model testing is the process of finding errors or inaccuracies within a simulation model.
Hence, with regard to verification and validation, the simulation quality describes the degree to which simulation model(s) and their coupling methods fulfill both aspects during model development and application.

In the urban intersection example described in section~\ref{sec_example}, simulation quality includes the following parts which are also shown on the left side of Fig.~\ref{fig:chapter_flow}:
\begin{itemize}
    \item[1:] Nanoscopic: quality of single atomic simulation models, e.g., pedestrian model, subunits of opponent vehicles,
    \item[2:] Microscopic: quality of coupling, e.g., synchronization, message format, actor availability,
    \item[3:] Macroscopic: quality of coupled system, e.g., opponent vehicle, simulation environment module.
\end{itemize}
There are several methods for quality assessment of simulation models.
An overview can be found in Riedmaier \textit{et al.} \cite{riedmaier2020framework}:
the focus is on a single atomic or coupled simulation model (1) and (3), and several quality criteria for validation are proposed, depending on the kind of simulation model (deterministic vs. non-deterministic) as well as the output characteristics (boolean, probabilistic, real-valued).
Additionally, unit tests for simulation models, partial simulations, comparison with real-world data, or fault-injections fall into category (1) and (3).
Another example is given by Frerichs \textit{et al.} \cite{frerichs2018quality}, where the simulation model of a steering system is tested.
In the context of simulation quality, the quality of atomic simulation models or units is called \textbf{nanoscopic} and the quality of coupled simulation models is called \textbf{macroscopic}.
Between atomic and coupled simulation model quality falls the quality of the coupling mechanism (2), which is called \textbf{microscopic} simulation quality.
The difference between (2) as opposed to (1) and (3) is the focus on either the unit behavior or the coupling between units.
(2) and (3) can lead to several iterations until the quality of the coupled units is assessed properly.
If (atomic) simulation models are coupled, we would ideally expect that by using validated coupling mechanisms, we could automatically assume that the coupled simulation models are of high quality.
For most applications, however, this is unfortunately not the case.
On its own, the coupling quality can be rated as good, but in an overall system with multiple simulation models, errors can still occur.
Examples for the quality of coupling are checking actor availability, synchronization between actors, or data availability in message protocols.
However, to our knowledge, there are no established standards yet.

Moreover, we assume that a general statement on the simulation quality can never be made and that measured quality is only an approximation of the actual quality.
The description of simulation quality might lead to the impression that tool qualification is done by a bottom-up approach.
However, it does not contradict a top-down approach 
and merely states on what levels tool qualification can take place.
Table~\ref{tab:smq} shows a few examples for simulation model testing. 
There are many methods for verification and validation of simulation models and the ISO 26262 provides a standard for tool qualification. 
However, it is hard to come by methods for coupling strategies of simulation models.

\begin{figure}[]
	\centering
	\includegraphics[width=\linewidth]{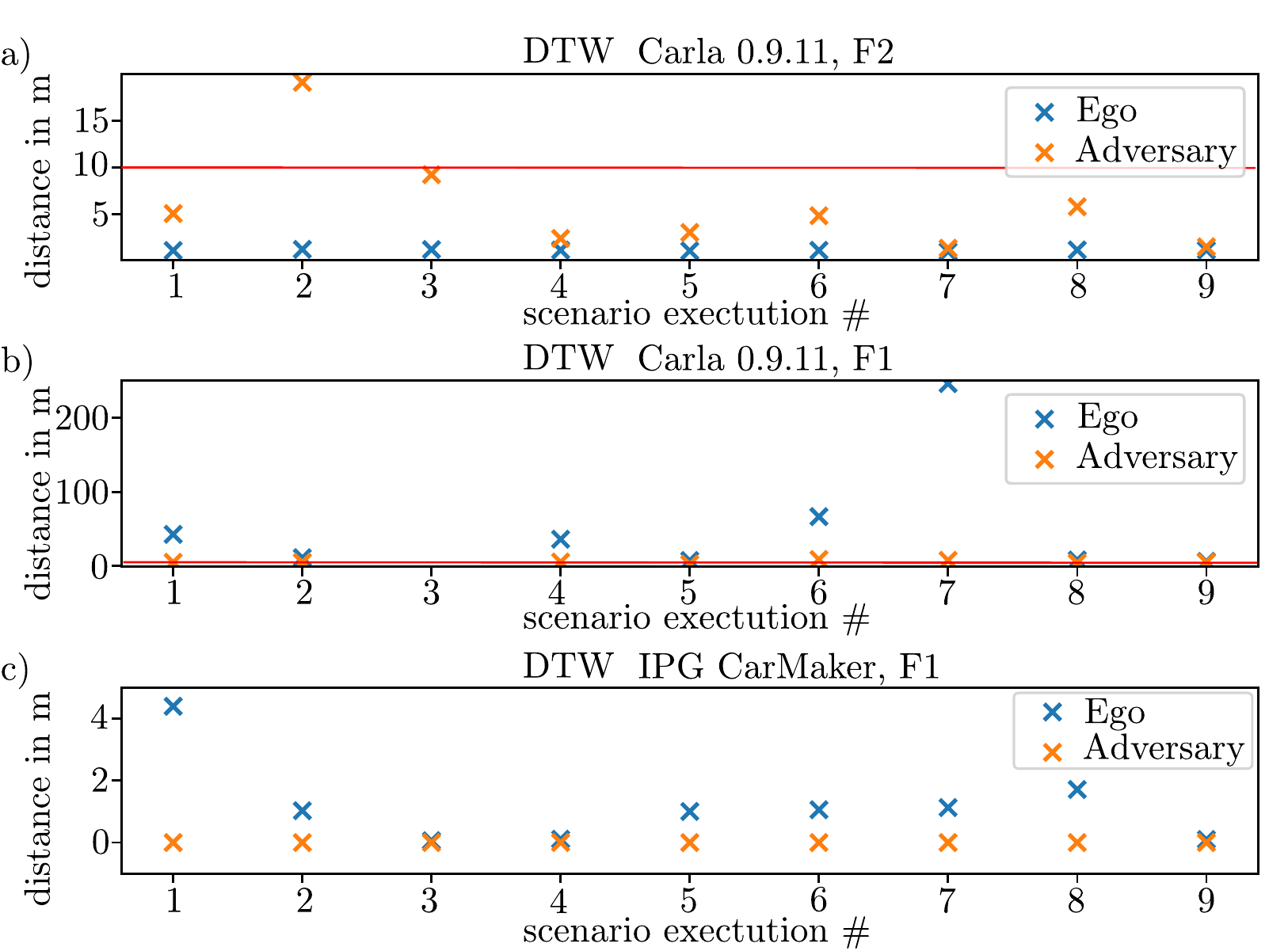}
	\caption{Dynamic Time Warping (DTW) comparison of different ego driving functions (F1, F2) and adversary vehicles in different simulation environments (Carla, CarMaker) for several executions of the same scenario description.}
	\label{fig:sim_quality_comp}
\end{figure}

For a better understanding, the example intersection scenario from section~\ref{sec_example} is used in different simulation tools.
In order to give a statement about the reliability of simulation results of different simulation environment setups and, therefore, a macroscopic quality, we compared execution results from each setup among several executions of the same scenario.
Two simulation environments were used: Carla 0.9.11 and CarMaker 8.0.2 from IPG combined with two different driving functions, called F1 and F2 respectively.
F1 is an external driving function connected via the open simulation interface (OSI) \cite{osi} and F2 the BasicAgent from Carla's PythonAPI \cite{Dosovitskiy17}.
The same intersection scenario was executed ten times for each setup where the first execution was used as reference point for following results.
The trajectories of the ego vehicle and an adversary vehicle were compared with the reference scenario's trajectories via dynamic time warping (DTW), a metric to get the distance between two time series \cite{ries2021trajectory}.
Ideally and to indicate deterministic results, the distance should be $\SI{0.0}{\meter}$ or close to $\SI{0.0}{\meter}$ for nearly identical simulation results.
We used $\SI{10.0}{\meter}$ as a threshold, which is indicated by the red lines in Fig.~\ref{fig:sim_quality_comp} a) and b); in c) no threshold is shown since all values lie beyond this threshold.
If a simulation consists of $2000$ steps this means that the average trajectory deviation per step should be smaller than $\SI{10/2000}{\meter} = \SI{0.005}{\meter}$.

Fig.~\ref{fig:sim_quality_comp} shows the results for all executions. 
The only actor with completely deterministic behavior is the adversary vehicle in Fig.~\ref{fig:sim_quality_comp} c), which is a trajectory follower and part of the CarMaker software.
The biggest DTW distances can be observed for F1 in Fig.~\ref{fig:sim_quality_comp} b), where scenario 3 has a value over 5000 for the ego and over 9000 for the adversary vehicle (both values are not shown since the scale would make it even more difficult to read other results). In this scenario, the ego vehicle came off course and drove a circle at the intersection, which results in a very different trajectory than the reference trajectory. Additionally, it influences the behavior of the opponent, e.g., an adversary vehicle might need to wait longer until it can pass the intersection if it is occupied by the ego vehicle.
There are several possible reasons why these distances are as high as in Fig.~\ref{fig:sim_quality_comp} and indicate a poor macroscopic simulation quality: the used driving functions are not reliable enough (low system under test quality, which is explained further in section~\ref{sec_sut_q}, coupling methods are of bad quality (low microscopic simulation quality), or non-deterministic physics model (if a deterministic model is needed for simulation a non-deterministic model indicates low nanoscopic simulation quality).
Fig.~\ref{fig:sim_quality_comp} a) shows a DTW value around $1.0$ for each F2 trajectory and since Carla 0.9.11. uses the non-deterministic physics model from its Unreal engine, it indicates that in the best case there are minor differences but overall results are still close to each other. 
The adversary vehicle has higher DTW values since the adversary vehicle reacts to the ego vehicle and, therefore, the differences from the ego actor is added and results in even higher trajectory deviations mostly in time.
Furthermore, Fig.~\ref{fig:sim_quality_comp} c) demonstrates that CarMaker's adversary vehicle is capable of deterministic behavior, however, F1 still shows unreliable results but all values are sill below the given threshold of $\SI{10.0}{\meter}$.
From these observations and the knowledge we have about the simulators, the results indicates further experiments are needed to evaluate the microscopic simulation quality (coupling of the driving function), where problems might come from the used PID controller or the message interface (OSI) as well as evaluation of the system under test (driving function F1) quality in all three levels of resolution.

This example shows, that the proposed taxonomy offers terms and terminology for a faster and clearer understanding and communication of areas of interest within the evaluation of simulation results.

\begin{table}
	\centering
	\caption{Simulation quality in literature}
	\setlength{\tabcolsep}{3pt}
	\begin{tabular}{p{43pt} p{133pt} p{34pt}}
		\toprule
		Level& 
		Description& 
		Publication \\
		\midrule
		\midrule
		Nanoscopic &
		simulation model V$\&$V methods survey &\cite{riedmaier2020framework}\\ \cmidrule{2-3}
		& steering system evaluation &\cite{frerichs2018quality}\\ \cmidrule{2-3}
		&scenario-based model evaluation&\cite{kuefler2017imitating}\\
		\midrule
		Microscopic & accuracy, effort, efficiency 
		&\cite{Gnther2017BeitragZC}
		\\
		\midrule
		Macroscopic &
		simulation model V$\&$V methods survey &\cite{riedmaier2020framework}\\ \cmidrule{2-3}
		& ISO 26262 Standard: tool qualification&\cite{iso26262}\\
		\bottomrule
	\end{tabular}
	\label{tab:smq}
\end{table}

\subsection{System Under Test Quality}
\label{sec_sut_q}

System under test quality evaluates the observable behavior and performance of a system under test with respect to the desired or intended functionality according to predefined requirements.
A special and widely used sub-class of system under test quality is safety quality, which evaluates how \textit{safe} a system under test can handle certain situations.
Safety can be functional safety according to ISO 26262 \cite{iso26262}, where it is described as ``absence of unreasonable risk due to hazards caused by malfunctioning behavior of Electrical/Electronic systems''.
A further approach is to assess safety in critical scenarios, where critical situations are derived from a prior criticality analysis \cite{neurohr2021criticality}.
A common example is a near-collision situation evaluated with metrics like the time-to-collision (TTC) metric \cite{hayward_near_1972}.
Assessing safety, e.g., functional safety, 
needs to follow defined development and test processes and tool qualification rules.
Whereas in the early stages of the development process, where proof of concepts and ideas are tested, performance might play a more significant role for developers than safety.
Additionally, it is important to note that different quality metrics can contradict each other: improving comfortable braking might also lead to more collisions in critical situations.

\begin{figure}[]
	\centering
	\includegraphics[width=\linewidth]{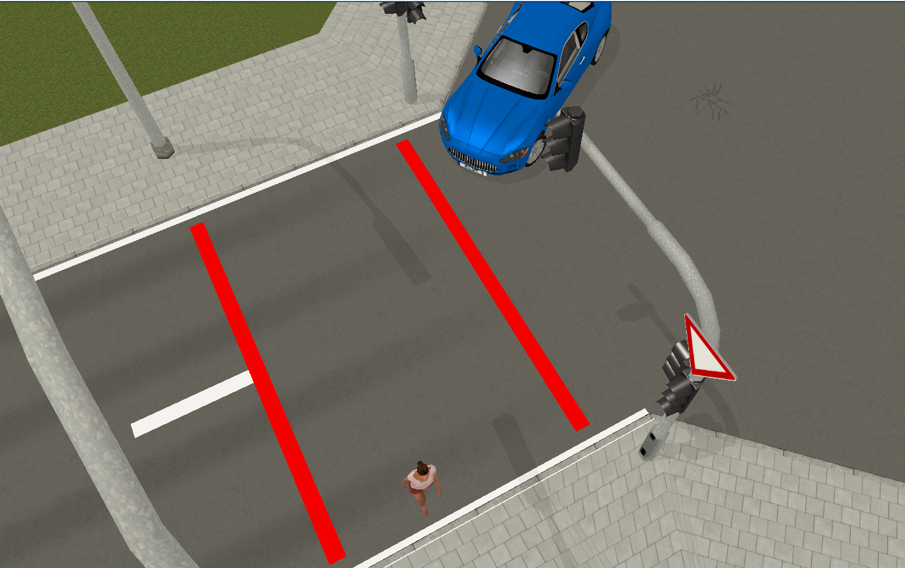}
	\caption{Example scenario with ego vehicle and pedestrian at an intersection as it is shown in dSpace \cite{dSpacesim}.}
	\label{fig:scenario_scene}
\end{figure}

In the urban intersection example described in section~\ref{sec_example}, the system under test quality shall investigate how well the driving function can avoid collisions in critical situations and might include the following quality metrics which are also shown on the top right side of Fig.~\ref{fig:chapter_flow}:
\begin{itemize}
    \item[4:] Nanoscopic: quality of system under test for one time step, e.g., headway or gap time,
    \item[5:] Microscopic: quality of system under test for a time interval, e.g., post-encroachment-time or encroachment-time,
    \item[6:] Macroscopic: quality of system under test for a set of time intervals, e.g., overall collision probability for functional scenario or ODD.
\end{itemize}
A scenario-based test approach is used, and functional and logical scenarios as well as their pass/fail criteria can be derived \cite{menzel2018scenarios}.
We propose one possible functional scenario: at an urban intersection, the ego vehicle shall turn right; by entering the right arm of the intersection, a pedestrian crosses the street.
To keep the logical scenario simple, only three variables were introduced that can vary throughout the derived concrete scenarios: the maximum speed $v_\text{max}$ allowed for the ego vehicle, the time $t_\text{cross}$ the pedestrian needs to cross the street, and the starting distance $d_\text{start}$ between ego vehicle and pedestrian, that has to be reached for the pedestrian to start crossing the intersection.
Concrete scenarios can then be obtained and executed when all variable ranges are defined and Fig.~\ref{fig:scenario_scene} shows an example scene from a simulation of a possible concrete scenario.
In our example, the derived concrete scenarios consist of all possible combinations of the three variables, where possible values for $v_\text{max}$ are from \SIrange{30}{58}{\km\per\hour} with step size \SI{2}{\km\per\hour}, $t_\text{cross}$ from \SIrange{5}{9}{\second} 
with step size \SI{1}{\second}, and $d_\text{start}$ from \SIrange{10}{24}{\meter} with step size \SI{2}{\meter}.
After deriving and concretizing scenarios, these concrete scenarios can be executed in a simulator and gained information can be summarized and combined to assess the system under test quality.

At first, quality at the time step level can be evaluated.
This step is associated with the matrix entry (4) in Fig.~\ref{fig:qualitymatrix} and Fig.~\ref{fig:chapter_flow}.
Information on this level can differ throughout time series, e.g., the distance between two traffic participants.
\Figure[t!](topskip=0pt, botskip=0pt, midskip=0pt)[width=0.99\textwidth]{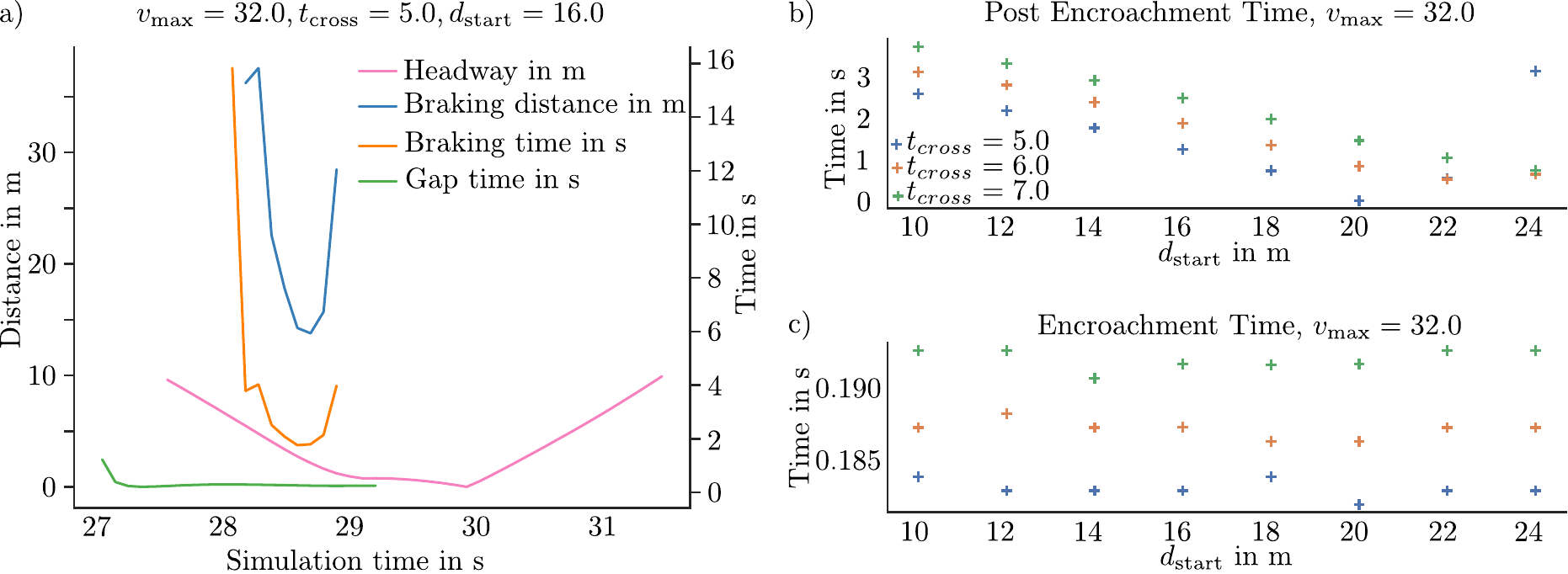}
{a) Example for nanoscopic metrics. b) + c) Examples for microscopic metrics.\label{fig:example_nano_micro}}
Fig.~\ref{fig:example_nano_micro}~a) shows the metric results calculated for four different quality metrics at the time step level of the intersection scenario with $v_\text{max}=\SI{32}{\km\per\hour}$, $t_\text{cross}=\SI{5}{\second}$, and $d_\text{start}=\SI{16}{\meter}$.
The used quality metrics are braking time, braking distance, gap time, and headway (distance) between the ego vehicle and the pedestrian during the simulation.
gap time describes the predicted distance in time between vehicle and pedestrian passing the intersection of their trajectories.
The fact that it is going towards \SI{0}{\second} means there is a near-collision situation or even a collision.
Gap time cannot be measured anymore when one actor passes the intersection of both trajectories and, therefore, the green gap time graph stops after about \SI{29.2}{\second} when the application period of gap time has passed.
The gaps in the graphs of braking time and distance show the system under test either stood still or tried to accelerate (where braking time and distance approach infinity) in between phases of braking since the application period condition for both metrics is a negative acceleration.
Additionally, threshold values can be defined for the considered quality metrics, e.g., $\text{Gap Time} > \SI{2}{\second}$. 
Another performance quality example could be to evaluate comfortable braking behavior.
Simulation results as depicted in Fig.~\ref{fig:example_nano_micro}~a)  are called \textbf{nanoscopic} system under test quality.

The next entry is information evaluation on a time interval level (5), where quality criteria and their results can be used to gain more information on a time interval, e.g., (partial) scenario.
A time interval can also be defined through an application period, e.g., if two traffic participants are within a certain distance and only if this condition is true other metrics are used.
Fig.~\ref{fig:example_nano_micro}~b)  and c) show the metric results of post-encroachment-time (PET) and  encroachment-time (ET), respectively and each value belongs to a concrete scenario derived from the logical scenario in the urban intersection example.
According to Allen \textit{et al.} \cite{Allen.1978},  post-encroachment-time is defined as the actual time gap between two traffic participants passing the intersection point or area of their trajectories.
Encroachment-time is the time an actor is occupying the intersection point or area and, therefore, describes the time it is exposed to a possible accident.
As shown in Fig.~\ref{fig:example_nano_micro}~b), encroachment-time results slightly increase with the time the pedestrian needs to cross the street ($t_{\text{cross}}$), but, as expected, the ego vehicle's starting distance and ego vehicle's speed have no impact as they are not related to the pedestrian's movement.
Therefore, in the urban intersection example the quality criteria are the quality metrics post-encroachment-time and  encroachment-time. 
The metric results  of these metrics can be evaluated with the example threshold values $\text{PET} > \SI{1.5}{\second}$ and $\text{ET} < \SI{5}{\second}$. 
For both metrics the application time is controlled by the time the actors pass the trajectory intersection.
Another possible quality metric is the smallest measured distance between two actors during one scenario.
Smallest measured distance is the microscopic version of the nanoscopic headway.
Simulation results of time intervals as depicted in Fig.~\ref{fig:example_nano_micro} are called \textbf{microscopic} system under test quality.

\begin{table}
	\centering
	\caption{System under test quality in literature}
	\setlength{\tabcolsep}{3pt}
	\begin{tabular}{p{45pt} p{115pt} p{45pt}}
		\toprule
		Level& 
		Description& 
		Publication \\
		\midrule
		\midrule
		Nanoscopic &
		Time-to-collision&\cite{hayward_near_1972}, \cite{de2017assessment}\\ \cmidrule{2-3}
		&Gap Time&\cite{Allen.1978}\\
		\midrule
		Microscopic &
		PET, ET & \cite{Allen.1978}\\ \cmidrule{2-3}
		&min. Time-to-collision&\cite{de2017assessment}\\ 
		\midrule
		Macroscopic&
		collision probability&\cite{de2017assessment}\\
		\bottomrule
	\end{tabular}
	\label{tab:sutq}
\end{table}

Quality evaluation on the next abstraction level is called \textbf{macroscopic} since it combines microscopic quality criteria of different time intervals, test series, or aggregated scenarios regarding a system under test.
This step is associated with entry (6).
A functional scenario can lead to different logical scenarios, e.g., similar situations on different maps, different types of pedestrians, e.g., children or handicapped with walking aids.
Additionally, one logical scenario can be implemented differently: actors can follow predefined trajectories or only be given goal positions they have to reach.
In the urban intersection example, results from scenarios derived from the functional scenario can be combined with other near-collision scenarios and evaluated, e.g., calculate overall collision probability in functional scenario or ODD. 
Table~\ref{tab:sutq} lists a few examples for system under test quality evaluation.
Elements for which system under test quality is measured can be highly automated driving functions, assisted driving functions, or even simulation models. For instance, the system under test is a simulation model, then nanoscopic, microscopic, and macroscopic system under test quality can be used to get nanoscopic simulation quality for this simulation model.

\subsection{Scenario Quality}
\label{sec_scenario_q}
\begin{figure}[]
	\centering
	\includegraphics[width=\linewidth]{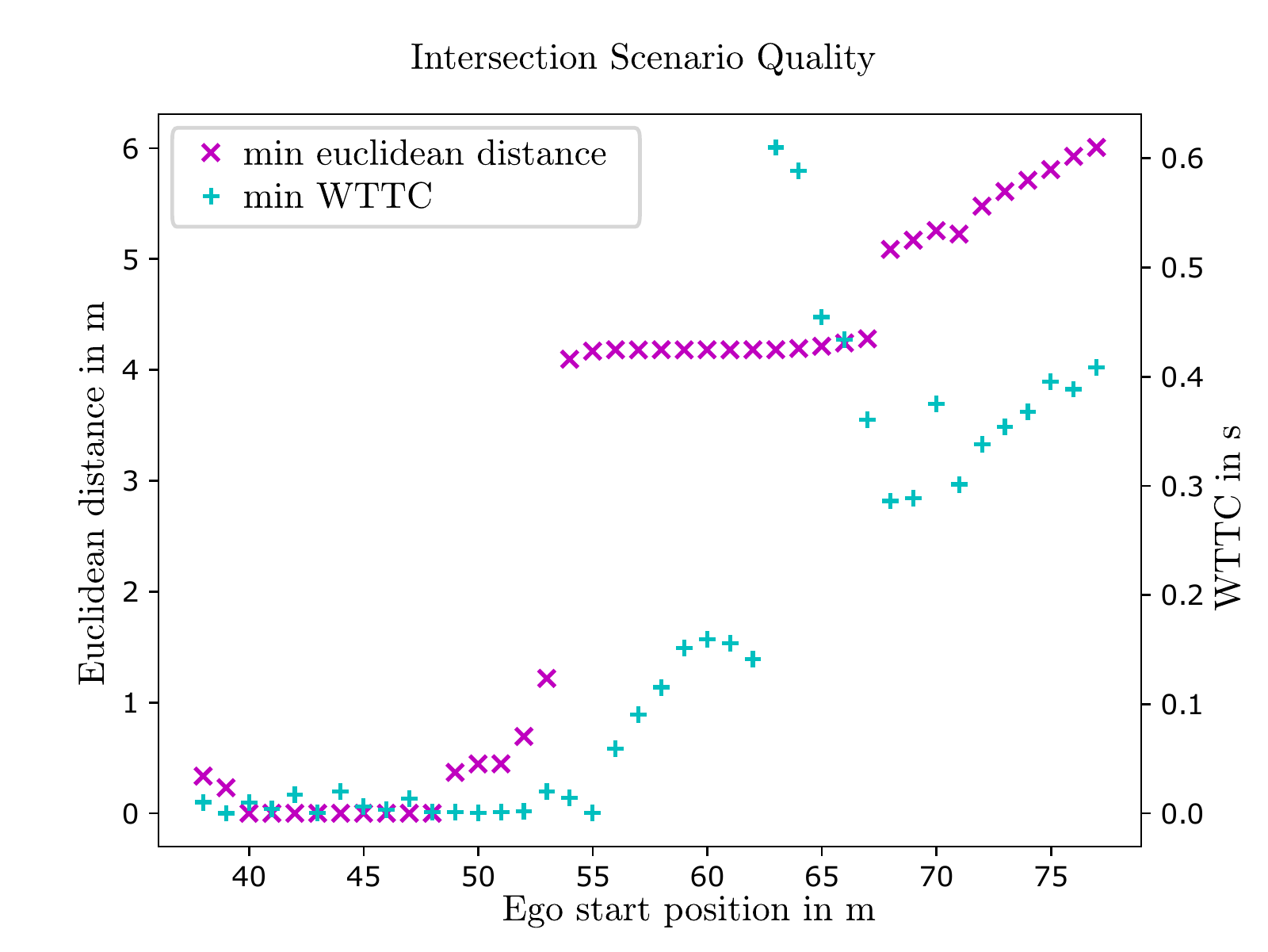}
	\caption{Euclidean distance and worst time-to-collision (WTTC) between ego vehicle and pedestrian as microscopic scenario quality.}
	\label{fig:scen_quality_comp}
\end{figure}

In some phases during the development and test process it is necessary to evaluate the quality of a scenario segment, a single scenario, or a set of scenarios, e.g., to determine the criticality of a situation regarding the criticality analysis of a system under test \cite{neurohr2021criticality}.
In this case, the quantified element is the scenario itself.
Scenario quality may have overlaps with system under test or traffic quality, and it depends on the viewpoint or use case which category it falls into.
For instance, the distance between two cars can be used to evaluate an adaptive cruise control system \cite{basantis2019standardized} but also to find new scenarios \cite{schiller2019multi}.
Another scenario quality worth mentioning is scenario coverage, which plays an essential role in overall test coverage.
Reducing mileage is one of the goals of scenario-based testing, and to do this, it is important to be able to make a statement about the coverage.
Scenario quality metrics can give evidence about the criticality within a scenario or its relevance for a test series.
The difference between system under test quality and scenario quality is the focus of the evaluation. 
In system under test quality, all metrics pertain to the system under test, whereas scenario quality focuses on various properties of a scenario, however, system under test quality can be part of it.
Quality criteria for scenarios can be a composition of different quality metrics or the same metrics taken from different traffic participants and evaluated together.

In the urban intersection example described in section~\ref{sec_example}, two goals could be to find scenarios with near-collision situations or scenarios with insufficient environmental information due to covert road users for further testing and for monitoring the quality of collision avoidance during different software revisions. 
Fig.~\ref{fig:scen_quality_comp} shows the minimum euclidean distance and worst time-to-collision (WTTC) \cite{7535468} between ego vehicle and the road crossing pedestrian.
In this case, the focus does not lie on the performance of the ego vehicle but on the microscopic scenario quality, the end value for an executed scenario.
The nanoscopic qualities for this scenario are the euclidean distances and worst time-to-collision that where measured for every time step during the simulation with their microscopic version of choosing the minimum value for a complete simulated scenario, respectively.

This example shows, that the proposed taxonomy helps to easily distinguish and communicate the difference in the area of interest. In this case, the performance of the ego vehicle is not the focus of the evaluated scenarios but the scenario itself. For instance, a set of scenarios gained this way can be used to evaluate changes that were made to the used driving function in a later revision.

All scenarios compared in Fig.~\ref{fig:scen_quality_comp} where the same, and only the x coordinate of the ego vehicle's starting position varied between $\SI{38}{\meter}$ and $\SI{78}{\meter}$.
Over the course of these scenarios it is possible to see that the worst time-to-collision starts with scenarios close to $\SI{0}{\second}$, but increases the further away the ego vehicle is set.
Regarding worst time-to-collision it could be interesting to do more simulations with a starting point between position $\SI{63}{\meter}$ and $\SI{64}{\meter}$ since there is a gap between the results, which often indicates a change in the order and course of the scenario and its participants.
The same holds true for starting positions $\SI{53}{\meter}$ and $\SI{54}{\meter}$ regarding the minimum euclidean distance.

Possible metrics for scenario quality, which are also shown on the bottom right side of Fig.~\ref{fig:chapter_flow} are:
\begin{itemize}
    \item[7:] Nanoscopic: quality of scenario segment, e.g., time-to-collision (time step), headway (time step), or traffic density (time step or time interval),
    \item[8:] Microscopic: quality of single scenario, e.g., min(time-to-collision) or min(headway),
    \item[9:] Macroscopic: quality of set of scenarios, e.g., variable coverage within this set.
\end{itemize}

\begin{table}
	\centering
	\caption{Scenario quality in literature}
	\setlength{\tabcolsep}{3pt}
	\begin{tabular}{p{45pt} p{95pt} p{65pt}}
		\toprule
		Level& 
		Description& 
		Publication \\
		\midrule
		\hline
		Nanoscopic&TTC&\cite{kluck2019performance}\\\cmidrule{2-3}
		&traffic quality& \cite{hallerbach2018simulation}\\\cmidrule{2-3}
		&WTTC&  \cite{7535468}\\
		\midrule
		Microscopic &
		scenario uniqueness & \cite{langner2018estimating}\\\cmidrule{2-3}
		&\makecell[l]{search-based techniques\\ }& \makecell[l]{\cite{schiller2019multi} (min. distance),\\ \cite{kluck2019performance} (min. TTC)} \\
		\midrule
		Macroscopic &coverage&\cite{hauer2019did}\\
		\bottomrule
	\end{tabular}
	\label{tab:sceq}
\end{table}
Example metrics as used in section~\ref{sec_sut_q} are possible on a \textbf{nanoscopic} (7) level and quantify the scenario quality for a scenario segment, i.e., time step, time interval or application time depending on conditions that are not fulfilled during the complete scenario. The next level of scenario quality is the \textbf{microscopic} (8) level, where the quality of a complete scenario is evaluated.
The last abstraction level is \textbf{macroscopic} scenario quality, where the aggregation of a set of scenarios is evaluated (9).
A common example for macroscopic scenario evaluation is coverage \cite{hauer2019did}.
Table~\ref{tab:sceq} lists literature about scenario quality evaluation.


\section{Conclusion and Future Work}
\label{sec:conclusion}
In this paper, we defined and delineated the concept of quality and important terms that are relevant during the simulation process.
We explained all relevant terms in section~\ref{sec:glossary} and gave an overview of their relationships to each other.
Based on these terms, we analyzed different domains of interest and different simulation resolution types and proposed a classification to assess quality for each aspect.
For this purpose, terms regarding resolution and established in other domains were introduced: macroscopic, microscopic, and nanoscopic.
This approach eases the evaluation process since it defines comparable aspects during the verification and validation process and clearly defines what level of information a quality criterion provides.
A taxonomy for three different quality classes was introduced, which was represented in a two-dimensional matrix.
This taxonomy shows that the classification is a way to help scientists to imminently understand and communicate the differences of areas of quality in the field of automotive simulation.

As long as simulation model quality is not good enough, simulation models have to be developed further until they can be used for testing (1)-(3).
Scenario quality decides if scenarios are interesting and relevant for a function at hand and can give a statement about coverage, etc., and can help to find new scenarios or better fitting ones (7)-(9).
Finally, system under test quality shows \textit{how good} a function is and if it needs more revisions during development or how it performs compared to others (4) - (6).
If other domains of interest arise, they can be easily added to the taxonomy.

Additionally, quality metrics for simulation models and driving functions can be used (among scenario-specific metrics) to evaluate scenario quality and scenarios can be used to test simulation models and driving functions.
However, this subjective method might lead to a cyclical dependence: system under test/simulation model quality metrics are used to find critical scenarios, and these, in turn, are then used to test the system under test/simulation model.
In our opinion, it is important to distinguish between scenario and system under test/simulation model quality to allow more objective metrics. 
Metrics that do not center around the performance of an ego vehicle, but on other features, e.g., assess the start set up of a scenario, can also give insight to the criticality within a scenario and could be used to define universal baseline sets of scenarios to ensure certain quality standards independent of driving functions or simulation models.

In the future, quality assessment and metrics can be related to this taxonomy for an easier understanding and classification, and if needed, new domains of interest can be added.
Moreover, new tools and standards are important to assess and compare quality throughout the development and test process.
However, simulation model verification and validation need further systematic approaches for a better quality evaluation in general.
In particular, the entries (2) and (9) in Fig.~\ref{fig:qualitymatrix} can be researched further to establish useful, well-defined, and safe methods to ensure overall simulation quality and in scenario quality more objective evaluation metrics in combination with existing ones are needed, e.g., to define baseline scenario sets for testing.


	\bibliographystyle{IEEEtran}
	\bibliography{paper}

\begin{thebibliography}{10}
\providecommand{\url}[1]{#1}
\csname url@samestyle\endcsname
\providecommand{\newblock}{\relax}
\providecommand{\bibinfo}[2]{#2}
\providecommand{\BIBentrySTDinterwordspacing}{\spaceskip=0pt\relax}
\providecommand{\BIBentryALTinterwordstretchfactor}{4}
\providecommand{\BIBentryALTinterwordspacing}{\spaceskip=\fontdimen2\font plus
\BIBentryALTinterwordstretchfactor\fontdimen3\font minus
  \fontdimen4\font\relax}
\providecommand{\BIBforeignlanguage}[2]{{%
\expandafter\ifx\csname l@#1\endcsname\relax
\typeout{** WARNING: IEEEtran.bst: No hyphenation pattern has been}%
\typeout{** loaded for the language `#1'. Using the pattern for}%
\typeout{** the default language instead.}%
\else
\language=\csname l@#1\endcsname
\fi
#2}}
\providecommand{\BIBdecl}{\relax}
\BIBdecl

\bibitem{Urbanization}
K.~K. Goldewijk, A.~Beusen, and P.~Janssen, ``Long-term dynamic modeling of
  global population and built-up area in a spatially explicit way: Hyde 3.1,''
  \emph{The Holocene}, vol.~20, no.~4, pp. 565--573, 2010,
  doi:\url{10.1177/0959683609356587}.

\bibitem{bach2017taxonomy}
J.~Bach, S.~Otten, and E.~Sax, ``A taxonomy and systematic approach for
  automotive system architectures-from functional chains to functional
  networks,'' in \emph{Int. Conf. on Vehicle Technology and Intelligent
  Transport Systems}, vol.~2.\hskip 1em plus 0.5em minus 0.4em\relax
  SCITEPRESS, 2017, pp. 90--101.

\bibitem{wood_safety_2019}
\BIBentryALTinterwordspacing
M.~Wood, P.~Robbel, M.~Maass, R.~D. Tebbens, M.~Meijs, M.~Harb, and
  P.~Schlicht, ``Safety first for automated driving,'' \emph{Aptiv, Audi, BMW,
  Baidu, Continental Teves, Daimler, FCA, HERE, Infineon Technologies, Intel,
  Volkswagen}, 2019. [Online]. Available:
  \url{https://www.daimler.com/documents/innovation/other/safety-first-for-automated-driving.pdf}
\BIBentrySTDinterwordspacing

\bibitem{dict2020}
C.~U. Press, ``Cambridge dictionary,'' \url{https://dictionary.cambridge.org/},
  2020, accessed: Oct. 21, 2020. [Online].

\bibitem{wachenfeld2016release}
W.~Wachenfeld and H.~Winner, ``The release of autonomous vehicles.''\hskip 1em
  plus 0.5em minus 0.4em\relax Springer, 2016, pp. 425--449.

\bibitem{pegasus_method}
{Pegasus Project}, ``{Pegasus Method - An Overview},''
  \url{https://www.pegasusprojekt.de/files/tmpl/Pegasus-Abschlussveranstaltung/PEGASUS-Gesamtmethode.pdf},
  2019, accessed: Apr. 16, 2020. [Online].

\bibitem{menzel2018scenarios}
T.~Menzel, G.~Bagschik, and M.~Maurer, ``Scenarios for development, test and
  validation of automated vehicles,'' in \emph{2018 IEEE Intelligent Vehicles
  Symp. (IV)}.\hskip 1em plus 0.5em minus 0.4em\relax IEEE, 2018, pp.
  1821--1827, doi: \url{10.1109/IVS.2018.8500406}.

\bibitem{Bode}
E.~B{\"o}de, M.~B{\"u}ker, U.~Eberle, M.~Fr{\"a}nzle, S.~Gerwinn, and
  B.~Kramer, ``Efficient splitting of test and simulation cases for the
  verification of highly automated driving functions,'' in \emph{Computer
  Safety, Reliability, and Security}, B.~Gallina, A.~Skavhaug, and F.~Bitsch,
  Eds.\hskip 1em plus 0.5em minus 0.4em\relax Cham: Springer International
  Publishing, 2018, pp. 139--153, doi: \url{10.1007/978-3-319-99130-6_10}.

\bibitem{king2019automated}
C.~King, L.~Ries, C.~Kober, C.~Wohlfahrt, and E.~Sax, ``Automated function
  assessment in driving scenarios,'' in \emph{2019 12th IEEE Conf. Software
  Testing, Validation and Verification (ICST)}.\hskip 1em plus 0.5em minus
  0.4em\relax IEEE, 2019, pp. 414--419, doi: \url{10.1109/ICST.2019.00050}.

\bibitem{bock2019advantageous}
F.~Bock, C.~Sippl, A.~Heinz, C.~Lauerz, and R.~German, ``Advantageous usage of
  textual domain-specific languages for scenario-driven development of
  automated driving functions,'' in \emph{2019 IEEE Int. Systems Conf.
  (SysCon)}.\hskip 1em plus 0.5em minus 0.4em\relax IEEE, 2019, pp. 1--8, doi:
  \url{10.1109/SYSCON.2019.8836912}.

\bibitem{ulbrich2015defining}
S.~Ulbrich, T.~Menzel, A.~Reschka, F.~Schuldt, and M.~Maurer, ``Defining and
  substantiating the terms scene, situation, and scenario for automated
  driving,'' in \emph{2015 IEEE 18th Int. Conf. Intelligent Transportation
  Systems}.\hskip 1em plus 0.5em minus 0.4em\relax IEEE, 2015, pp. 982--988,
  doi: \url{10.1109/ITSC.2015.164}.

\bibitem{bach2016model}
J.~Bach, S.~Otten, and E.~Sax, ``Model based scenario specification for
  development and test of automated driving functions,'' in \emph{2016 IEEE
  Intelligent Vehicles Symp. (IV)}.\hskip 1em plus 0.5em minus 0.4em\relax
  IEEE, 2016, pp. 1149--1155, doi: \url{10.1109/IVS.2016.7535534}.

\bibitem{openscenario}
{ASAM OpenSCENARIO}, ``{ASAM OpenSCENARIO},''
  \url{https://www.asam.net/standards/detail/openscenario/}, 2020, accessed:
  Apr. 21, 2021. [Online].

\bibitem{kramer2020}
B.~Kramer, C.~Neurohr, M.~B{\"u}ker, E.~B{\"o}de, M.~Fr{\"a}nzle, and W.~Damm,
  ``Identification and quantification of hazardous scenarios for automated
  driving,'' in \emph{Model-Based Safety and Assessment}.\hskip 1em plus 0.5em
  minus 0.4em\relax Springer, 2020, pp. 163--178, doi:
  \url{10.1007/978-3-030-58920-2_11}.

\bibitem{sae2018surface}
SAE, ``{J3016 - SURFACE VEHICLE RECOMMENDED PRACTICE - Taxonomy and Definitions
  for Terms Related to Driving Automation Systems for On-Road Motor
  Vehicles},'' \url{https://www.sae.org/standards/content/j3016{\_}201806/},
  2018, accessed: Oct. 20, 2020. [Online].

\bibitem{schutt2020sceml}
B.~Sch{\"u}tt, T.~Braun, S.~Otten, and E.~Sax, ``Sceml: a graphical modeling
  framework for scenario-based testing of autonomous vehicles,'' in \emph{Proc.
  23rd ACM/IEEE Int. Conf. Model Driven Engineering Languages and Systems},
  2020, pp. 114--120, doi: \url{10.1145/3365438.3410933}.

\bibitem{Steimle2020TerminologyEnglish}
M.~Steimle, T.~Menzel, and M.~Maurer, ``{Towards a Consistent Terminology for
  Scenario-Based Development and Test Approaches for Automated Vehicles: A
  Proposal for a Structuring Framework, a Basic Vocabulary, and its
  Application},'' \url{https://arxiv.org/abs/2104.09097}, 2021, accessed: Apr.
  21, 2021. [Online].

\bibitem{VTDsim}
\BIBentryALTinterwordspacing
{MSC Software}, ``Virtual test drive (vtd) - complete tool-chain for driving
  simulation,'' accessed: Oct. 08, 2021. [Online]. [Online]. Available:
  \url{https://www.mscsoftware.com/product/virtual-test-drive}
\BIBentrySTDinterwordspacing

\bibitem{dSpacesim}
\BIBentryALTinterwordspacing
{dSpace}, ``dspace - automotive simulation models (asm).'' [Online]. Available:
  \url{https://www.dspace.com/en/inc/home/products/sw/automotive{\_}
  simulation{\_}models.cfm}
\BIBentrySTDinterwordspacing

\bibitem{IPGsim}
\BIBentryALTinterwordspacing
{IPG}, ``Ipg - automotive gmbh - everything about virtual test driving,''
  accessed: Oct. 08, 2021. [Online]. [Online]. Available:
  \url{https://ipg-automotive.com/}
\BIBentrySTDinterwordspacing

\bibitem{Dosovitskiy17}
A.~Dosovitskiy, G.~Ros, F.~Codevilla, A.~Lopez, and V.~Koltun, ``{CARLA}: {An}
  open urban driving simulator,'' in \emph{Proceedings of the 1st Annual
  Conference on Robot Learning}, 2017, pp. 1--16.

\bibitem{SUMO2018}
\BIBentryALTinterwordspacing
P.~A. Lopez, M.~Behrisch, L.~Bieker-Walz, J.~Erdmann, Y.-P. Fl{\"o}tter{\"o}d,
  R.~Hilbrich, L.~L{\"u}cken, J.~Rummel, P.~Wagner, and E.~Wie{\ss}ner,
  ``Microscopic traffic simulation using sumo,'' in \emph{The 21st IEEE
  International Conference on Intelligent Transportation Systems}.\hskip 1em
  plus 0.5em minus 0.4em\relax IEEE, 2018. [Online]. Available:
  \url{https://elib.dlr.de/124092/}
\BIBentrySTDinterwordspacing

\bibitem{openPass}
\BIBentryALTinterwordspacing
{openPASS}, ``openpass - target objectives,'' accessed: Oct. 20, 2021.
  [Online]. [Online]. Available:
  \url{https://openpass.eclipse.org/target_objectives/}
\BIBentrySTDinterwordspacing

\bibitem{Vissim}
\BIBentryALTinterwordspacing
{PTV Group}, ``Traffic simulation software - ptv vissim - ptv group,''
  accessed: Oct. 08, 2021. [Online]. [Online]. Available:
  \url{https://www.ptvgroup.com/en/solutions/products/ptv-vissim/}
\BIBentrySTDinterwordspacing

\bibitem{esmini}
\BIBentryALTinterwordspacing
{esmini}, ``esmini: a basic openscenario player,'' accessed: Oct. 08, 2021.
  [Online]. [Online]. Available: \url{https://github.com/esmini/esmini}
\BIBentrySTDinterwordspacing

\bibitem{5706287}
IEEE, ``Ieee recommended practice for distributed simulation engineering and
  execution process (dseep),'' \emph{IEEE Std 1730-2010 (Revision of IEEE Std
  1516.3-2003)}, pp. 1--79, 2011, doi: \url{10.1109/IEEESTD.2011.5706287}.

\bibitem{durak2020safety}
U.~Durak, A.~D'Ambrogio, and P.~Bocciarelli, ``Safety-critical simulation
  engineering,'' in \emph{Proceedings of the 2020 Summer Simulation
  Conference}, 2020, pp. 1--12, doi: \url{10.5555/3427510.3427535}.

\bibitem{iso26262}
ISO, ``{ISO26262: Road vehicles -- Functional safety},'' {2018}.

\bibitem{ni2006framework}
D.~Ni, ``A framework for new generation transportation simulation,'' in
  \emph{Proc. 2006 Winter Simulation Conf.}\hskip 1em plus 0.5em minus
  0.4em\relax IEEE, 2006, pp. 1508--1514, doi: \url{10.1109/WSC.2006.322920}.

\bibitem{schiller2019multi}
M.~Schiller, M.~Dupius, D.~Krajzewicz, A.~Kern, and A.~Knoll,
  ``Multi-resolution traffic simulation for large-scale high-fidelity
  evaluation of vanet applications,'' in \emph{Simulating Urban Traffic
  Scenarios}.\hskip 1em plus 0.5em minus 0.4em\relax Springer, 2019, pp.
  17--36, doi: \url{10.1007/978-3-319-33616-9_2}.

\bibitem{kups1274}
\BIBentryALTinterwordspacing
N.~G. Eissfeldt, ``Vehicle-based modelling of traffic . theory and application
  to environmental impact modelling,'' Ph.D. dissertation, Universit{\"a}t zu
  K{\"o}ln, 2004. [Online]. Available: \url{https://kups.ub.uni-koeln.de/1274/}
\BIBentrySTDinterwordspacing

\bibitem{viehof2018research}
M.~Viehof and H.~Winner, ``Research methodology for a new validation concept in
  vehicle dynamics,'' \emph{Automotive and Engine Technology}, vol.~3, no. 1-2,
  pp. 21--27, 2018, doi: \url{10.1007/s41104-018-0024-1}.

\bibitem{rosenberger2019towards}
P.~Rosenberger, J.~T. Wendler, M.~F. Holder, C.~Linnhoff, M.~Bergh{\"o}fer,
  H.~Winner, and M.~Maurer, ``Towards a generally accepted validation
  methodology for sensor models-challenges, metrics, and first results,'' 2019.

\bibitem{riedmaier2020framework}
S.~Riedmaier, B.~Danquah, B.~Schick, and F.~Diermeyer, ``Unified framework and
  survey for model verification, validation and uncertainty quantification,''
  \emph{Archives of Computational Methods in Engineering}, pp. 1886--1784,
  2020, doi: \url{10.1007/s11831-020-09473-7}.

\bibitem{westhofen2021criticality}
L.~Westhofen, C.~Neurohr, T.~Koopmann, M.~Butz, B.~Schütt, F.~Utesch,
  B.~Kramer, C.~Gutenkunst, and E.~Böde, ``Criticality metrics for automated
  driving: a review and suitability analysis of the state of the art,'' 2021.

\bibitem{gettman2003surrogate}
D.~Gettman and L.~Head, ``Surrogate safety measures from traffic simulation
  models,'' \emph{Transportation Research Record}, vol. 1840, no.~1, pp.
  104--115, 2003, doi: \url{10.3141/1840-12}.

\bibitem{abeysirigoonawardena2019generating}
Y.~Abeysirigoonawardena, F.~Shkurti, and G.~Dudek, ``Generating adversarial
  driving scenarios in high-fidelity simulators,'' in \emph{2019 Int. Conf. on
  Robotics and Automation (ICRA)}.\hskip 1em plus 0.5em minus 0.4em\relax IEEE,
  2019, pp. 8271--8277, doi: \url{10.1109/ICRA.2019.8793740}.

\bibitem{hallerbach2018simulation}
S.~Hallerbach, Y.~Xia, U.~Eberle, and F.~Koester, ``Simulation-based
  identification of critical scenarios for cooperative and automated
  vehicles,'' \emph{SAE International Journal of Connected and Automated
  Vehicles}, vol.~1, no. 2018-01-1066, pp. 93--106, 2018, doi:
  \url{10.4271/2018-01-1066}.

\bibitem{junietz2019microscopic}
P.~M. Junietz, ``Microscopic and macroscopic risk metrics for the safety
  validation of automated driving,'' Ph.D. dissertation, Technische
  Universit{\"a}t, 2019.

\bibitem{Allen.1978}
{Allen, Brian, L.}, B.~T. Shin, and {Cooper, Peter, J.}, ``{Analysis of Traffic
  Conflicts and Collisions},'' \emph{{Transportation Research Record}}, vol.
  667, pp. 67--74, 1978.

\bibitem{balci1998verification}
O.~Balci, ``Verification, validation, and testing,'' \emph{Handbook of
  simulation}, vol.~10, no.~8, pp. 335--393, 1998.

\bibitem{frerichs2018quality}
D.~Frerichs and M.~Borsdorf, ``Quality for vehicle system simulation,'' in
  \emph{VDI-Kongress" SIMVEC-Simulation und Erprobung in der
  Fahrzeugentwicklung}, 2018.

\bibitem{osi}
\BIBentryALTinterwordspacing
{ASAM e.V.}, ``Welcome to open simulation interface’s documentation!''
  accessed: Oct. 08, 2021. [Online]. [Online]. Available:
  \url{https://opensimulationinterface.github.io/osi-documentation/index.html}
\BIBentrySTDinterwordspacing

\bibitem{ries2021trajectory}
L.~Ries, P.~Rigoll, T.~Braun, T.~Schulik, J.~Daube, and E.~Sax,
  ``Trajectory-based clustering of real-world urban driving sequences with
  multiple traffic objects,'' in \emph{The 24th IEEE International Conference
  on Intelligent Transportation Systems}.\hskip 1em plus 0.5em minus
  0.4em\relax IEEE, 2021.

\bibitem{kuefler2017imitating}
A.~Kuefler, J.~Morton, T.~Wheeler, and M.~Kochenderfer, ``Imitating driver
  behavior with generative adversarial networks,'' in \emph{2017 IEEE
  Intelligent Vehicles Symp. (IV)}.\hskip 1em plus 0.5em minus 0.4em\relax
  IEEE, 2017, pp. 204--211, doi: \url{10.1109/IVS.2017.7995721}.

\bibitem{Gnther2017BeitragZC}
F.~G{\"u}nther, ``Beitrag zur {C}o-{S}imulation in der
  {G}esamtsystementwicklung des {K}raftfahrzeugs,'' 2017.

\bibitem{neurohr2021criticality}
C.~Neurohr, L.~Westhofen, M.~Butz, M.~Bollmann, U.~Eberle, and R.~Galbas,
  ``Criticality analysis for the verification and validation of automated
  vehicles,'' \emph{IEEE Access}, 2021, doi: \url{10.1109/ACCESS.2021.3053159}.

\bibitem{hayward_near_1972}
J.~C. Hayward, ``Near miss determination through use of a scale of danger,'' in
  \emph{Unknown}, 1972, publisher: Pennsylvania State University University
  Park.

\bibitem{de2017assessment}
E.~de~Gelder and J.-P. Paardekooper, ``Assessment of automated driving systems
  using real-life scenarios,'' in \emph{2017 IEEE Intelligent Vehicles
  Symposium (IV)}.\hskip 1em plus 0.5em minus 0.4em\relax IEEE, 2017, pp.
  589--594, doi: \url{10.1109/IVS.2017.7995782}.

\bibitem{basantis2019standardized}
A.~Basantis, L.~Harwood, Z.~Doerzaph, and L.~Neurauter, ``Standardized
  performance evaluation of vehicles with automated capabilities,'' 2019, doi:
  \url{10.15787/VTT1/D946JJ}.

\bibitem{7535468}
W.~Wachenfeld, P.~Junietz, R.~Wenzel, and H.~Winner, ``The
  worst-time-to-collision metric for situation identification,'' in \emph{2016
  IEEE Intelligent Vehicles Symposium (IV)}, 2016, pp. 729--734, doi:
  \url{10.5555/3427510.3427535}.

\bibitem{kluck2019performance}
F.~Kl{\"u}ck, M.~Zimmermann, F.~Wotawa, and M.~Nica, ``Performance comparison
  of two search-based testing strategies for adas system validation,'' in
  \emph{IFIP Int. Conf. on Testing Software and Systems}.\hskip 1em plus 0.5em
  minus 0.4em\relax Springer, 2019, pp. 140--156, doi:
  \url{10.1007/978-3-030-31280-0_9}.

\bibitem{langner2018estimating}
J.~Langner, J.~Bach, L.~Ries, S.~Otten, M.~Holz{\"a}pfel, and E.~Sax,
  ``Estimating the uniqueness of test scenarios derived from recorded
  real-world-driving-data using autoencoders,'' in \emph{2018 IEEE Intelligent
  Vehicles Symposium (IV)}.\hskip 1em plus 0.5em minus 0.4em\relax IEEE, 2018,
  pp. 1860--1866, doi: \url{10.1109/IVS.2018.8500464}.

\bibitem{hauer2019did}
F.~Hauer, T.~Schmidt, B.~Holzm{\"u}ller, and A.~Pretschner, ``Did we test all
  scenarios for automated and autonomous driving systems?'' in \emph{2019 IEEE
  Intelligent Transportation Systems Conf. (ITSC)}.\hskip 1em plus 0.5em minus
  0.4em\relax IEEE, 2019, pp. 2950--2955, doi: \url{10.1109/ITSC.2019.8917326}.

\end{thebibliography}
	\balance
	
	\begin{IEEEbiography}[{\includegraphics[width=1in,height=1.25in,clip,keepaspectratio]{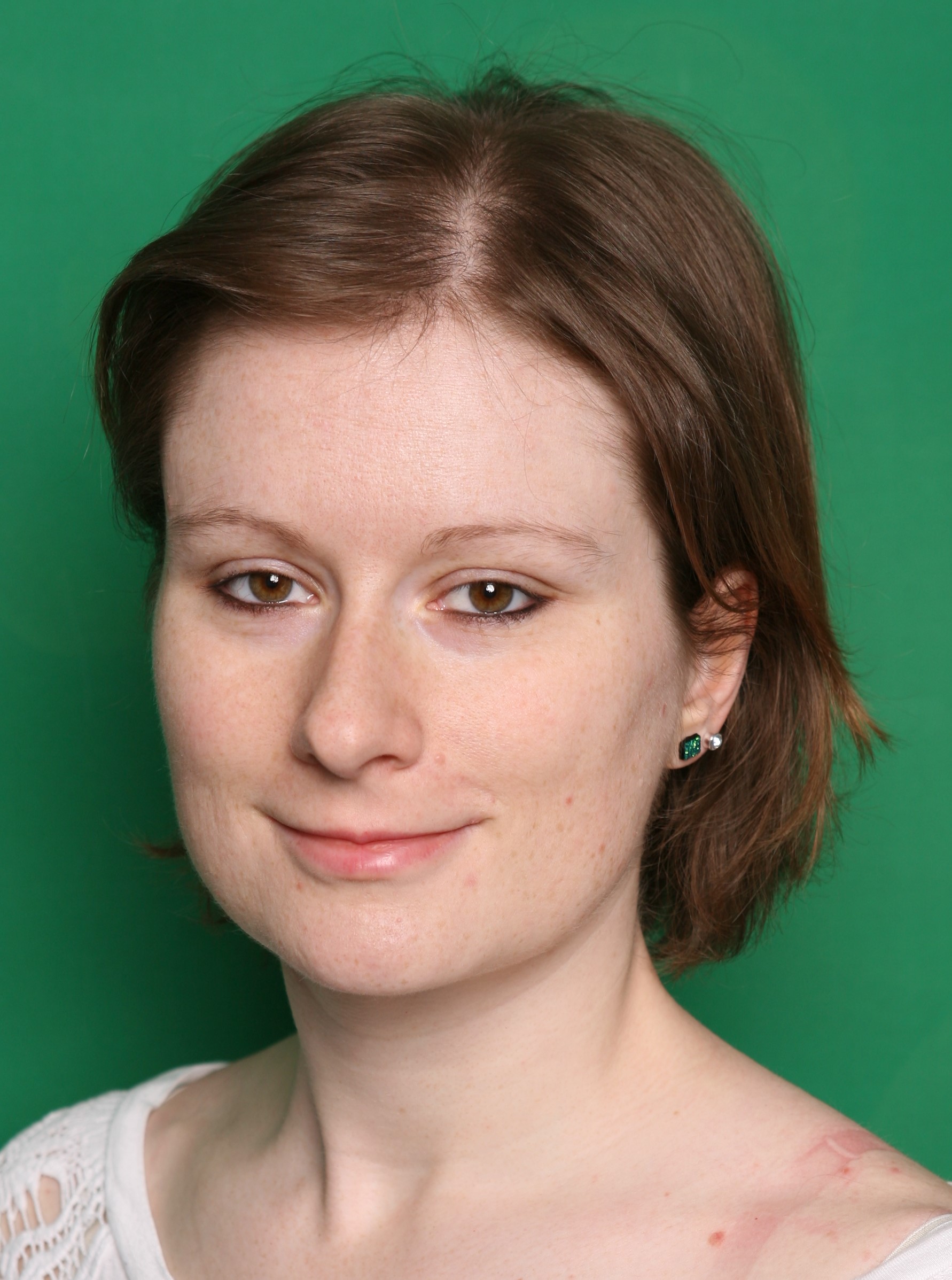}}]{Barbara Schütt} received a B.Sc. and an M.Sc. degrees in computer science from Technische Universität Berlin in 2012 and 2015, respectively. Before starting as a researcher and Ph.D. student at FZI Research Center for Information Technology in 2019, she was a researcher at Fraunhofer Institute for Production Systems and Design Technology in the department of robotics and automation and worked as a software developer in the series development at Carmeq/Volkswagen in the field of situation analysis for ADAS. The main focus of her research is on quality metrics and scenario evaluation.
	\end{IEEEbiography}
	
	\begin{IEEEbiography}[{\includegraphics[width=1in,height=1.25in,clip,keepaspectratio]{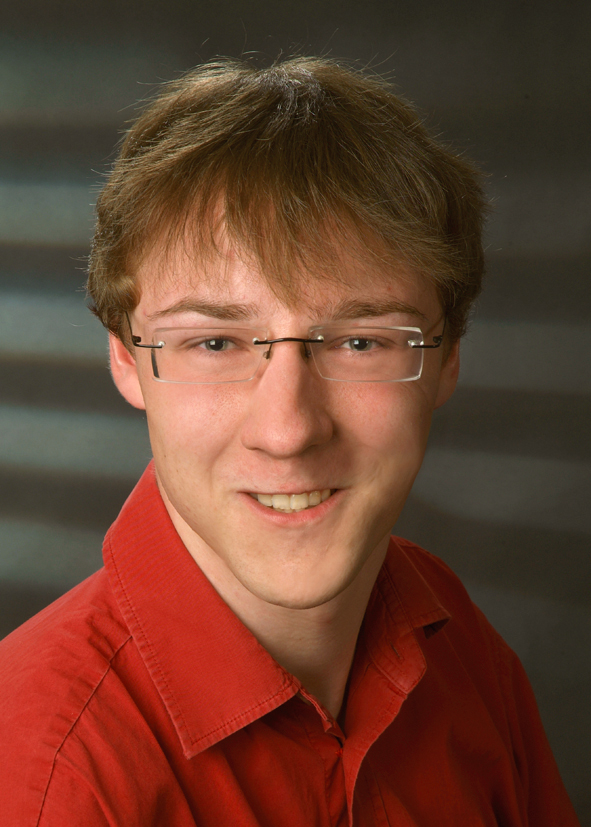}}]{Markus Steimle} received a B.Eng. degree in electrical engineering and information technology from the Landshut University of Applied Sciences, Landshut, Germany, in 2014.
    In 2016, he received an M.Sc. degree in automotive software engineering from the Technical University of Munich, Munich, Germany.
    Since 2016, he has been a Ph.D. student at the Institute of Control Engineering of Technische Universität Braunschweig, Braunschweig, Germany. 
    His main research interests are scenario-based verification and validation of automated vehicles, focusing on the use of simulative testing methods.
\end{IEEEbiography}
	
	\begin{IEEEbiography}[{\includegraphics[width=1in,height=1.25in,clip,keepaspectratio]{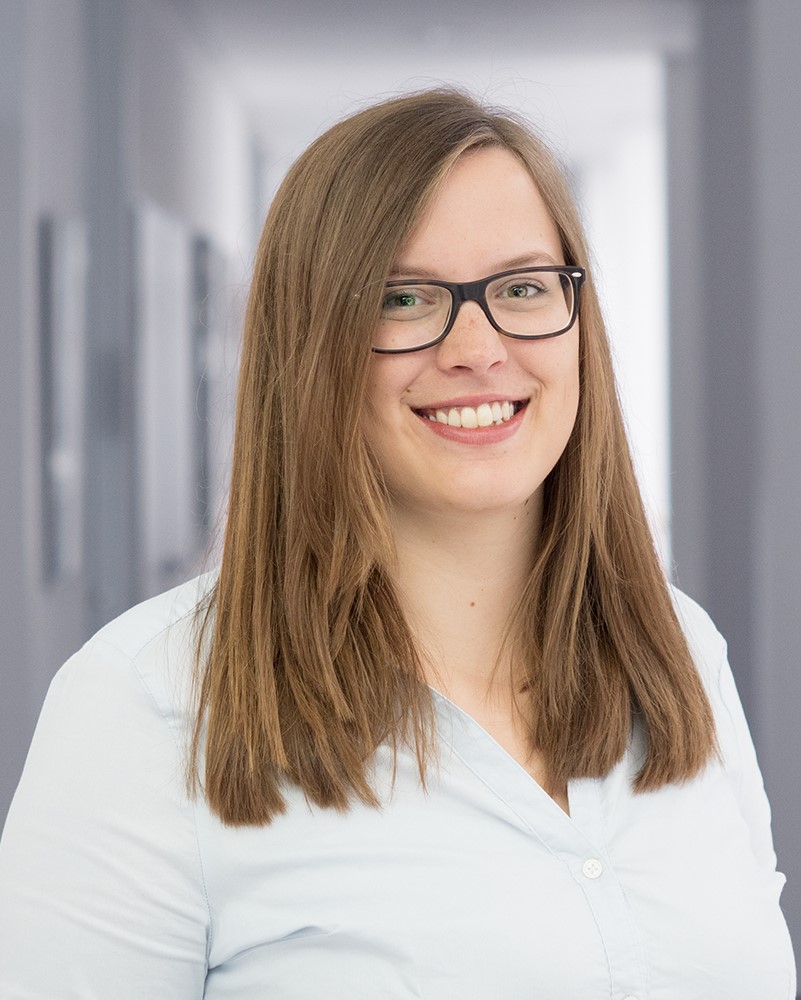}}]{Birte Kramer} received a B.Sc. and an M.Sc. degree in mathematics from the Carl von Ossietzky University in Oldenburg in 2015 and 2017. Since 2017 she is a researcher and Ph.D candidate at OFFIS e.V. where she is working in the area of scenario-based verification and validation of automated vehicles. Her main research interest concerns validity and quality criteria for simulations.
	\end{IEEEbiography}

	\begin{IEEEbiography}[{\includegraphics[width=1in,height=1.25in,clip,keepaspectratio]{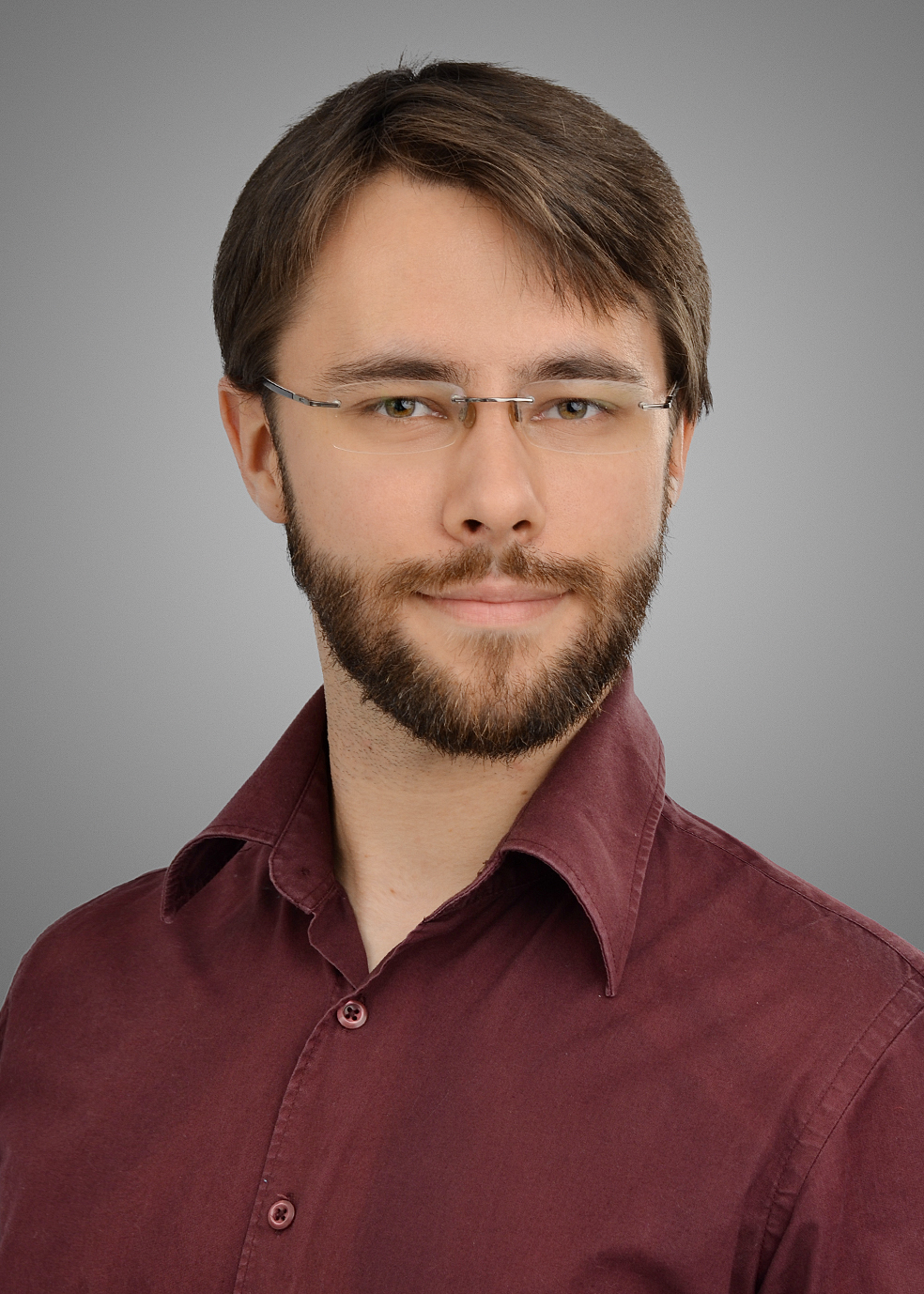}}]{Danny Behnecke} studied Computer and Communications Systems Engineering at Technische Universität Braunschweig (M. \, Sc. 2017). Since 2017 he is a research assistant at the Institute of Transportation Systems of the German Aerospace Center. His main research topics are simulation system architecture and validation of simulation execution.
	\end{IEEEbiography}

	\begin{IEEEbiography}[{\includegraphics[width=1in,height=1.25in,clip,keepaspectratio]{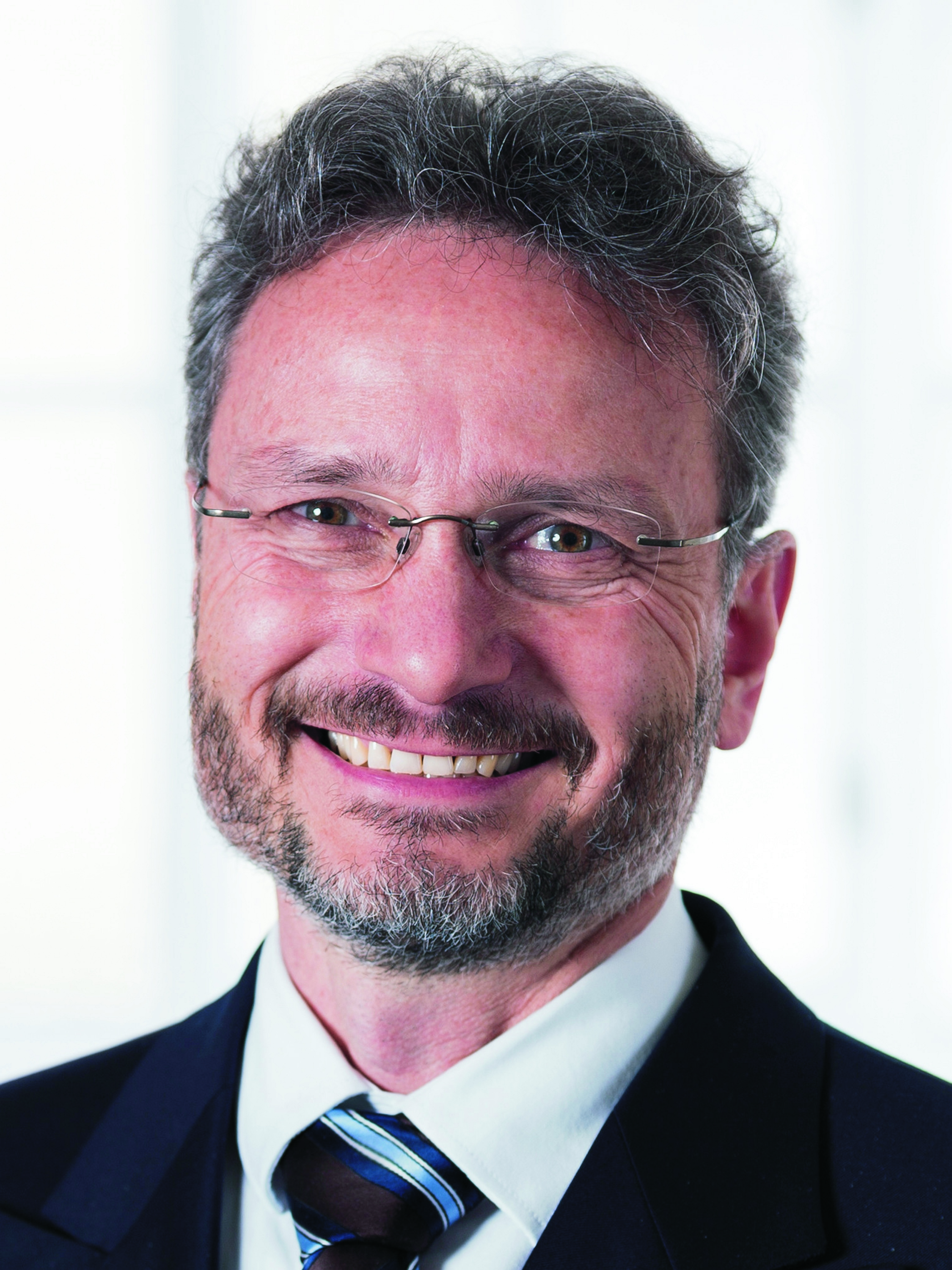}}]{Eric Sax} Prof. Dr.-Ing. Eric Sax is head of the Institute of Information Processing Technology (www.itiv.kit.edu) at the Karlsruher Institute of Technology. In addition, he is director at the Forschungszentrum Informatik (www.fzi.de) and the so called Hector School, the Technology Business School of KIT. His main topics of research, together with currently more than 50 PhD employees, are processes, methods and tools in systems engineering, focussing the automotive industry.
	\end{IEEEbiography}

\EOD	
\end{document}